\documentclass [11pt]{article}

\ifx\Red\relax\def\Red#1{#1}\fi
\def\Red#1{#1}

\def\textfraction{0.1}

\usepackage{epsfig}
\usepackage{amssymb}
\usepackage{verbatim}
\usepackage{delarray}
\usepackage{graphics}
\usepackage{epic}

\newcommand{\ket}[1]{{|#1\rangle}}

\newcommand{\inner}[2]{\langle{#1},{#2}\rangle}

\newcommand{\B}{{\mathcal{B}}}
\newcommand{\CC}{{\mathcal{C}}}

\newcommand{\G}{{\mathcal{G}}}
\newcommand{\HH}{{\mathcal{H}}}

\newcommand{\SSS}{{\mathcal{S}}}

\newcommand{\C}{{\mathbb{C}}}
\newcommand{\F}{{\mathbb{F}}}

\newcommand{\Z}{{\mathbb{Z}}}
\newcommand{\zerob}{{\mathbf 0}}
\newcommand{\ab}{{\mathbf a}}
\newcommand{\bb}{{\mathbf b}}
\newcommand{\cb}{{\mathbf c}}
\newcommand{\eb}{{\mathbf e}}
\newcommand{\gb}{{\mathbf g}}
\newcommand{\hb}{{\mathbf h}}

\renewcommand{\sb}{{\mathbf s}}
\newcommand{\tb}{{\mathbf t}}
\newcommand{\ub}{{\mathbf u}}

\newcommand{\Ab}{{\mathbf A}}
\newcommand{\Bb}{{\mathbf B}}
\newcommand{\Eb}{{\mathbf E}}
\newcommand{\Gb}{{\mathbf G}}
\newcommand{\Ib}{{\mathbf I}}
\newcommand{\Sb}{{\mathbf S}}

\newcommand{\omegabar}{\overline{\omega}}

\newcommand{\ie}{{\em i.e., }}
\newcommand{\eg}{{\em e.g., }}

\newcommand{\etal}{\emph{et al.\ }}

\newcommand{\Tr}{\mathrm{Tr~}}

\newcommand{\openbox}{\leavevmode
     \hbox to.77778em{%
     \hfil\vrule
     \vbox to.675em{\hrule width.6em\vfil\hrule}%
     \vrule\hfil}}
\newcommand{\proofname}{Proof}

\newcommand{\qed}{\hspace*{1cm}\hspace*{\fill}\openbox}

\newtheorem{theorem}{Theorem}

\title{Convolutional and Tail-Biting \\ Quantum
Error-Correcting Codes}
\author{G. David Forney, Jr.\footnote{Laboratory for
Information and  Decision Systems,
Massachusetts Institute of Technology,
Cambridge, MA 02139. E-mail: \texttt{forneyd@comcast.net}.},
Markus Grassl\footnote{Institut f\"ur Algorithmen und Kognitive
    Systeme, Universit\"at Karlsruhe (TH), 76128 Karlsruhe,
    Germany. E-mail: \texttt{grassl@ira.uka.de}.},
and
Saikat Guha\footnote{Research Laboratory of Electronics,
Massachusetts Institute of Technology,
Cambridge, MA 02139.
E-mail: \texttt{saikat@mit.edu}.}}

\begin{document}
\renewcommand{\textfraction}{0}

\date{}
\maketitle
\thispagestyle{empty}
\begin{abstract}

Rate-$(n-2)/n$ unrestricted and CSS-type quantum convolutional codes
with up to \Red{$4096$} states and minimum distances up to \Red{10} are
constructed  as stabilizer codes from classical self-orthogonal rate-$1/n$
$\F_4$-linear and binary linear convolutional codes, respectively.  These
codes generally have higher rate and less decoding complexity than
comparable quantum block codes or previous quantum convolutional codes.
Rate-$(n-2)/n$ block stabilizer codes with the same rate and
error-correction capability and essentially the same decoding complexity
are derived from these convolutional codes via tail-biting.

\end{abstract}
\normalsize


\section{Introduction}

Quantum error-correcting codes (QECCs) protect quantum
states from unwanted perturbations, allowing
the implementation of robust quantum computing and communication
systems.

The first breakthrough in this field was Shor's demonstration in 1995
via a $9$-qubit single-error-correcting code \cite{S95} that quantum
error-correction was even possible, which was not obvious
\emph{a priori}.  Shortly thereafter, a more efficient $7$-qubit
single-error-correcting code was found by Steane \cite{S96} and by
Calderbank and Shor \cite{CS96}.  This code was merely the first of a
class of quantum codes based on classical binary error-correcting codes,
which we call \emph{CSS-type} codes. Later in 1996, an even more
efficient $5$-qubit single-error-correcting code was found by Bennett
\etal \cite{BDSW96} and by Laflamme \etal \cite{LMPZ96}.

Soon thereafter, a general theory of \emph{stabilizer codes} was
developed \cite{CRSS97, G98, NC00, Presk}, which includes the above
codes as particular cases, and indeed essentially all QECCs developed to
date. The stabilizer formalism, which we review below, has the virtue
of reducing the QECC problem to pure mathematics, and thus allowing
non-physicists to contribute to the field.  In particular, it
shows how to convert certain classical $\F_4$-linear and binary
error-correcting codes to QECCs \cite{CRSS98}.

In this paper, we systematically develop quantum convolutional codes
(QCCs) using the same general principles.  We focus on rate-$1/n$ codes,
where we can exhibit some simple and attractive codes as examples, but
our construction principles are general.

Practical classical communication systems have mostly used
convolutional codes rather than block codes, because convolutional codes
are generally superior in terms of their performance-complexity
tradeoff.  In quantum coding, it is still too early to say which
characteristics of QECCs will turn out to be the most important.
However, we find that quantum convolutional codes compare favorably with
quantum block codes in the following ways:
\begin{itemize}
\item Code rate.  In general, QCCs require fewer encoded qubits to
protect the same number of information qubits than comparable block
codes.  For example, our rate-$1/3$ single-error-correcting QCCs are
comparable to the $5$-qubit and $7$-qubit single-error-correcting block
codes
mentioned above, but have higher code rate.
\item Decoding complexity.\footnote{In this paper, ``complexity" will
always mean decoding complexity;  we do not consider complexity issues at
the quantum gate level.  Moreover, the decoding computation after the
measurement of syndromes is  entirely classical.}  In general, QCCs have
simpler decoding algorithms.  For example, we present extremely simple
decoding algorithms for our single-error-correcting QCCs.
\item Performance.  In general, QCCs have a superior tradeoff between
performance and complexity.  For example, our  rate-1/3
single-error-correcting QCCs have comparable error probability to the
5-qubit and $7$-qubit block codes, even though they have higher rate and
simpler decoders.
\end{itemize}

One possible drawback of QCCs is their lack of a natural block
structure.  Previous authors have proposed terminating QCCs to yield
block codes with the same error-correction capability, at the cost of
reduced rate  (sometimes without recognizing that a
terminated convolutional stabilizer code may not be a valid
(\ie self-orthogonal) block stabilizer code;  see Section III-C.)  We
propose instead to construct quantum
\emph{tail-biting}
codes (QTBCs), which are block stabilizer codes that retain the same code
rate, error correction capability, and decoding algorithms as the QCCs from
which they are derived, provided that the block length is large enough.
We exhibit families of rate-$1/n$ QTBCs with attractive
performance-complexity tradeoffs.

Surprisingly, no one previously seems to have constructed QCCs that are
clearly superior to quantum block codes by a straightforward extension of
the stabilizer formalism. Chau \cite{C98, C99} proposed two 
``quantum convolutional codes:" the first involves a one-to-one
convolutional sequence transformation followed by a quantum block code, has
the same performance and complexity as the block code, and thus is arguably
not really ``convolutional;" the second generalizes Shor's concatenated
construction to convolutional codes, and yields a single-error-correcting
rate-1/4 QCC that corrects 1 error in every 8 qubits, comparable to a $[8,
2, 3]$ shortened quantum Hamming code (see Section II.E below).  Ollivier
and Tillich
\cite{OT03, OT04} constructed a rate-1/5 single-error-correcting QCC using
the stabilizer formalism, but  unfortunately their example QCC does not
improve on the comparable $5$-qubit block code in either performance or
complexity.  
However, they do address various gate-level implementation
issues that we do not address in this paper.  (For further results on
quantum convolutional encoders, see \cite{GR06}.)  More recently, Almeida
and Palazzo
\cite{AP04} have proposed rate-1/4, -1/3 and -2/4 convolutional codes using
a Shor-type concatenated construction;  these codes appear to be much more
complicated than ours to decode.

\pagebreak
In Section II, we briefly review  stabilizer
codes, particularly $\F_4$-linear and CSS-type codes.  In Section III,
we show how to construct quantum convolutional and tail-biting codes
from classical self-orthogonal $\F_4$-linear and binary convolutional
codes.  We give examples of simple rate-1/3
single-error-correcting $\F_4$-linear and CSS-type QCCs and QTBCS, and
their decoding algorithms.   In Section IV, we briefly
summarize the algebraic structure theory of rate-$1/n$ linear
shift-invariant convolutional codes and their orthogonal codes, and
specify the relevant symmetries of these codes in the QECC context. 
In Section V, we tabulate rate-$1/n$
single-error-correcting codes
and the corresponding tail-biting codes.  In Section VI, we present the
best rate-1/3 codes with state space sizes up to 2048 and minimum
distances up to 10,  with their corresponding tail-biting codes. 

\section{Review of stabilizer codes}

In this section we review the stabilizer formalism, originally developed
by Calderbank \emph{et al.\ }\cite{CRSS97} and Gottesman \cite{G98}, in
order to fix nomenclature and notation.  We focus especially on
$\F_4$-linear stabilizer codes \cite{CRSS98} and CSS-type codes
\cite{CS96, S96}.

\subsection{Qubits and Pauli matrices}

A \emph{qubit} is a quantum system whose
Hilbert space $\HH$ is two-dimensional.

Given a basis for $\HH$, a basic set of unitary
Hermitian operators on $\HH$ is the set $\Pi = \{I, X, Y, Z\}$ of
\emph{Pauli matrices}, defined by
$$
I = \left[\begin{array}{cc} 1 & 0 \\ 0 & 1 \end{array}\right],  \quad
X = \left[\begin{array}{cc} 0 & 1 \\ 1 & 0 \end{array}\right],  \quad
Y = \left[\begin{array}{cc} 0 & -i \\ i & 0 \end{array}\right],  \quad
Z = \left[\begin{array}{cc} 1 & 0 \\ 0 & -1 \end{array}\right].
$$
The multiplication table of these matrices is evidently as follows:
$$
\begin{array}{|c|cccc|}  \hline
\times  & I & X & Y & Z \\  \hline
I & I & X & Y & Z \\
X & X & I & iZ & -iY \\
Y & Y & -iZ & I & iX \\
Z & Z & iY & -iX & I \\ \hline
\end{array}
$$
Their commutation properties may therefore be summarized as
follows: if $A$ and $B$ are two Pauli matrices, then $AB = BA$ if $A$
or $B$ is the identity or if $A = B$;  otherwise $AB = -BA$.

The set $\Pi$ is not a multiplicative group, because
it is not closed under multiplication.  However, let us consider
instead the set $[\Pi] = \{[A] \mid A \in \Pi\}$ of equivalence classes
of Pauli matrices defined by $[A] = \{\beta A \mid  \beta \in \C,
|\beta| = 1\}$;  \ie $B$ is equivalent to $A \in \Pi$ if $B =
\beta A$, where $\beta$ is a unit-magnitude complex
number.\footnote{Previous authors restrict $\beta$ to $\{\pm 1, \pm
i\}$, which suffices to make $[\Pi]$ a group;  however, allowing
$\beta$ to range over all unit-magnitude complex numbers is
more natural physically.}
   From the multiplication table of $\Pi$, multiplication in
$[\Pi]$ is well defined and commutative, since $[A][B] = [B][A] = [AB] =
[BA]$.  The multiplication table of  $[\Pi]$ is
$$
\begin{array}{|c|cccc|} \hline
\times  & [I] & [X] & [Y] & [Z] \\  \hline
[I] & [I] & [X] & [Y] & [Z] \\
[0pt] [X] & [X] & [I] & [Z] & [Y] \\
[0pt] [Y] & [Y] & [Z] & [I] & [X] \\
[0pt] [Z] & [Z] & [Y] & [X] & [I] \\ \hline
\end{array}
$$
Thus $[\Pi]$ forms a commutative
(abelian) multiplicative group, which we will call the
\emph{projective Pauli group}.

By inspection, the projective Pauli group $[\Pi]$ is isomorphic
to the group $(\Z_2)^2 = \{00, 01, 10, 11\}$ of binary 2-tuples, whose
addition table is
$$
\begin{array}{|c|cccc|} \hline
+  & 00 & 10 & 11 & 01 \\  \hline
00 & 00 & 10 & 11 & 01 \\
10 & 10 & 00 & 01 & 11 \\
11 & 11 & 01 & 00 & 10 \\
01 & 01 & 11 & 10 & 00 \\ \hline
\end{array}
$$
Alternatively, the projective Pauli group is isomorphic to the
additive group of the quaternary field $\F_4 = \{0, 1, \omega,
\overline{\omega}\}$, whose addition table is
$$
\begin{array}{|c|cccc|}  \hline
+  & 0 & \omega & 1 & \overline{\omega} \\  \hline
0 & 0 & \omega & 1 & \overline{\omega} \\
\omega & \omega & 0 & \overline{\omega} & 1 \\
1 & 1 & \overline{\omega} & 0 & \omega \\
\overline{\omega} & \overline{\omega} & 1 & \omega & 0 \\ \hline
\end{array}
$$
In short, $0 + a = a$, $a + a = 0$ (so subtraction is the same as
addition), and $1 + \omega + \overline{\omega} = 0$.

These tables have been arranged to suggest that the elements of
$[\Pi]$, or their representatives in $\Pi$, may be labeled by elements
of $(\Z_2)^2$ or of $\F_4$ according to the following correspondences:
$$
\begin{array}{|c|c|c|} \hline
\Pi & (\Z_2)^2 & \F_4 \\  \hline
I & 00 & 0 \\
X & 10 & \omega \\
Y & 11 & 1 \\
Z & 01 & \overline{\omega} \\ \hline
\end{array}
$$
Then the \emph{label maps} $\ell:  [\Pi] \to (\Z_2)^2$ and $L: [\Pi] \to
\F_4$ that are defined by these correspondences are isomorphisms;
\ie
$$
\ell([A]) + \ell([B]) = \ell([AB]);  \qquad
L([A]) + L([B]) = L([AB]).
$$

By a slight abuse of notation, we may apply the label maps $\ell$ and
$L$ to $\Pi$, or to any matrices in the equivalence classes $[A] \in
[\Pi]$.
By a further slight abuse of notation, we may pass between two-bit and
quaternary labels via label maps $\ell:  \F_4 \to
(\Z_2)^2$ and $L: (\Z_2)^2 \to \F_4$.

The two label bits in $\ell(A)$, namely $\ell_1(A)$ and
$\ell_2(A)$, represent a bit flip and a phase
flip, respectively, since $X$ (or any $\beta X$, $|\beta| = 1$) is
a bit flip operator,
$Z$ is a phase flip operator, and $Y = iXZ$ is a combination of a bit
flip and a phase flip.

Finally, we may use the quaternary labels to characterize the
commutation
properties of Pauli matrices.    The \emph{traces} of the elements
$\{0,
1, \omega, \overline{\omega}\}$ of $\F_4$ are defined as
$\{0, 0, 1, 1\}$, and their \emph{conjugates} are defined as $\{0, 1,
\overline{\omega},  \omega\}$.
The \emph{Hermitian inner product} of two elements $a, b \in \F_4$ is
defined as $\inner{a}{b} = a^\dagger b \in \F_4$, where ``$^\dagger$"
denotes conjugation.  The \emph{trace inner product} is defined as
$\Tr \inner{a}{b} \in \F_2$.  Thus the multiplication,
Hermitian inner product, and trace inner product tables of $\F_4$ are
\pagebreak
$$
\begin{array}{|c|cccc|}  \hline
\times  & 0 & 1 & \omega & \overline{\omega} \\  \hline
0 & 0 & 0 & 0 & 0 \\
1 & 0 & 1 & \omega & \overline{\omega} \\
\omega & 0 & \omega & \overline{\omega} & 1 \\
\overline{\omega} & 0 & \overline{\omega} & 1 & \omega \\ \hline
\end{array}
\qquad
\begin{array}{|c|cccc|}  \hline
\inner{\,}{}  & 0 & 1 & \omega & \overline{\omega} \\  \hline
0 & 0 & 0 & 0 & 0 \\
1 & 0 & 1 & \omega & \overline{\omega} \\
\omega & 0 & \overline{\omega} & 1 & \omega \\
\overline{\omega} & 0 & \omega & \overline{\omega} & 1 \\ \hline
\end{array}
\qquad
\begin{array}{|c|cccc|}  \hline
\Tr\inner{\,}{}  & 0 & 1 & \omega & \overline{\omega} \\  \hline
0 & 0 & 0 & 0 & 0 \\
1 & 0 & 0 & 1 & 1 \\
\omega & 0 & 1 & 0 & 1 \\
\overline{\omega} & 0 & 1 & 1 & 0 \\ \hline
\end{array}
$$
Comparing the trace inner product table of $\F_4$ with the
multiplication
table of
$\Pi$, we see that  two matrices $A, B \in \Pi$ commute if
$\Tr \inner{L(A)}{L(B)} = 0$, and anticommute  if
$\Tr \inner{L(A)}{L(B)} = 1$.
In other words,
$$
AB = (-1)^{\Tr \inner{L(A)}{L(B)}} BA.
$$

A  corresponding inner product may be defined over $(\Z_2)^2$ (a
``symplectic" or ``twisted" inner product) such that a similar
result is obtained;  however, we will have no need for such a binary
inner product in this paper.

\subsection{Multi-qubit systems and Pauli $n$-tuples}

An \emph{$n$-qubit system} is a quantum system whose Hilbert space $\HH$
is the tensor product of $n$ two-dimensional spaces, and is thus
$2^n$-dimensional.

A Pauli $n$-tuple $\Ab = A_1\otimes A_2\otimes
\cdots\otimes A_n$ is a tensor product of $n$
Pauli matrices $A_i, 1 \le i \le n$, that act separately on each
of the $n$ qubits. The set of all $4^n$ Pauli $n$-tuples will be denoted
by $\Pi^n$.

    The product of two Pauli $n$-tuples is the componentwise
product of its elements:
$$
\Ab \Bb = (A_1 B_1)\otimes(A_2 B_2)\otimes
\cdots\otimes (A_n B_n).
$$
Consequently the product is a Pauli $n$-tuple up to phase.  If we again
define equivalence classes of Pauli $n$-tuples up to phase by $[\Ab] =
\{\beta \Ab \mid
\beta \in \C, |\beta| = 1\}$, then we obtain a well-defined product
$$
[\Ab] [\Bb] = [A_1 B_1]\otimes[A_2 B_2]\otimes \cdots
\otimes[A_n B_n] = [\Ab \Bb].
$$

Thus the set of $4^n$ equivalence classes of Pauli $n$-tuples
is a commutative multiplicative group $[\Pi^n]$ which is isomorphic to
$((\Z_2)^2)^n$. We may thus label the elements of $[\Pi^n]$ by binary
$2n$-tuples in
$((\Z_2)^2)^n$, or by quaternary $n$-tuples in $(\F_4)^n$, by extending
the label maps $\ell$ and $L$ of the previous subsection.  The
resulting label maps remain isomorphisms;  \ie
$$
\ell([\Ab]) + \ell([\Bb]) = \ell([\Ab \Bb]);  \qquad
L([\Ab]) + L([\Bb]) = L([\Ab\Bb]).
$$

Finally, the Hermitian inner product over $(\F_4)^n$ is defined
by the componentwise sum $\inner{\ab}{\bb} = \sum_i a_i^\dagger b_i.$
The trace inner product over $(\F_4)^n$ is therefore also a
componentwise sum:
$$
\Tr \inner{L(\Ab)}{L(\Bb)} = \Tr \left(\sum_{i=1}^n
L(A_i)^\dagger L(B_i)\right)  = \sum_{i=1}^n \Tr L(A_i)^\dagger L(B_i).
$$
Thus the single-qubit Pauli matrix commutation relation may be extended
to Pauli $n$-tuples:
$$\Bb\Ab = (-1)^{\Tr\inner{L(\Ab)}{L(\Bb)}} \Ab\Bb.$$

\subsection{Stabilizer and normalizer codes}

Within the stabilizer formalism, an $[n,k]$ \emph{stabilizer code} is
defined by a set of $n-k$ \emph{independent} (\ie none of the
generators is equivalent to a product of the others, up to phase)
commuting Pauli $n$-tuples $\G = \{\Gb_j, 1 \le j \le n-k\}$, where 
$0 \le k \le n$. The code subspace is the common eigenspace $\SSS$ of the
generators $\Gb_j \in \G$ such that the eigenvalue of each $\Gb_j \in \G$ is
+1;  \ie $\SSS$ is the subspace of $\HH$ that is \emph{stabilized} by $\G$.

The \emph{stabilizer group} $S \subset [\Pi^n]$ is defined as the set
of all of equivalence classes in $[\Pi^n]$ of all $2^{n-k}$ products
of the $n-k$ generators $\Gb_j \in \G$. By the independence condition,
the equivalence classes of these $2^{n-k}$ products are distinct.

The \emph{binary stabilizer label code} $\ell(S)$ is then the image of
$S$ under the binary label map.  Since $\ell: S \to \ell(S)$ must be
an isomorphism, $\ell(S)$ must be the classical $(2n, n-k)$ binary
linear block code that is generated by the $n-k$ linearly
independent generators $\ell(\Gb_j)$.

Similarly, the \emph{quaternary stabilizer label code} $L(S)$ is the
image of $S$ under the quaternary label map.
Evidently $L(S)$ is a group under addition that is isomorphic to
$\ell(S) \cong (\Z_2)^{n-k}$.

In order that all generators $\Gb_j \in \G$ commute, the trace inner
product of the quaternary labels of any two generators must be 0;  \ie
$L(S)$ must be self-orthogonal under the trace inner product.    This
holds if and only if the generators of $L(S)$ are
self-orthogonal and mutually orthogonal under the trace inner product.

We will focus on the case in which $L(S)$ is actually $\F_4$-linear;
\ie
closed under multiplication by scalars in $\F_4$.  $L(S)$ is then a
classical $(n, (n-k)/2)$ linear block code over $\F_4$.  This implies
that the integer $n-k$ must be even, and that $L(S)$ is the set of all
$\F_4$-linear combinations of $(n-k)/2$ independent generators $\gb_i, 1
\le i \le (n-k)/2$.

Equivalently, $L(S)$ is the set of all binary
linear combinations of the $n-k$ generators
$\omega \gb_i$ and $\overline{\omega} \gb_i$, $1 \le i \le (n-k)/2$.
This implies that the set $\G$ of generators $\Gb_j$ of $S$ may be
taken to be the $n-k$ inverse images in $\Pi^n$ of this set of
$(n-k)/2$ generator pairs under the inverse quaternary label map.
(Note that when inverting label maps defined on $\Pi^n$, we take
the inverse as the unique element of $\Pi^n$ with the given label; \ie
we fix the phase of the inverse image.)

Moreover, if $L(S)$ is $\F_4$-linear, then $L(S)$ must be
self-orthogonal
under the Hermitian inner product, not just the trace inner product,
since $\Tr \inner{\alpha \gb}{\hb} = \Tr \alpha \inner{\gb}{\hb}$ is
equal to 0 for $\alpha = 1, \omega, \overline{\omega}$ if and only if
$\inner{\gb}{\hb} = 0$.

The \emph{quaternary normalizer label code} is defined as the
orthogonal code $L(S)^\perp$ to $L(S)$ with respect to the trace inner
product;  \ie the set of all elements of $(\F_4)^n$ whose trace inner
product with all elements of $L(S)$ is zero. If
$L(S)$ is
$\F_4$-linear, then, by the same argument as above, the orthogonal code
$L(S)^\perp$ under the trace inner product must be equal to the orthogonal
code
$L(S)^\perp$ under the Hermitian inner product.  Thus $L(S)^\perp$ is a
classical $(n, (n+k)/2)$ $\F_4$-linear block code.

The inverse image of $L(S)^\perp$ under the inverse quaternary label map
is called the \emph{projective normalizer group} $N(S) \subseteq
[\Pi^n]$, or simply the ``normalizer group," a commutative subgroup of
$[\Pi^n]$ of size
$2^{n+k}$.  In other words, $L(N(S)) = L(S)^\perp$, so the normalizer
group is the set of equivalence classes of all Pauli $n$-tuples that
commute with all elements of the stabilizer group $S$.  Evidently $S$ is
a subgroup of $N(S)$, and the stabilizer label code $L(S)$ is a subcode
of $L(N(S))$.

\subsection{Quantum error correction}

We now explain briefly how an $[n,k]$ stabilizer code may be used to
encode the state of a $k$-qubit system into that of an $n$-qubit system,
and then to correct a certain set $\Sigma$ of error patterns.

\pagebreak
Let $\HH$ denote the $2^n$-dimensional Hilbert space of the $n$-qubit
system.  The $n-k$ independent commuting Pauli $n$-tuple generators
$\Gb_j$ of the stabilizer group $S$ each have eigenvalues $\{\pm 1\}$,
and have two corresponding orthogonal eigenspaces of dimension
$2^{n-1}$.  It is straightforward to show that since the $n-k$
generators
are independent and commuting, the stabilizer group $S$ that they
generate has a set of $2^{n-k}$ orthogonal
eigenspaces, each of dimension $2^k$, corresponding to the $2^{n-k}$
possible combinations of $\{\pm 1\}$ eigenvalues of the generators.

The $2^k$-dimensional eigenspace for which all
$n-k$ generator eigenvalues equal $+1$ (\ie the subspace stabilized by
$S$) is defined as the code subspace $\SSS \subseteq \HH$.
For encoding, first the Hilbert space of a $k$-qubit quantum system is
embedded into ${\cal H}$, \eg by initializing the remaining $n-k$ qubits
to a fixed state.  Then a unitary transformation maps this
$2^k$-dimensional subspace into $\SSS$.

The object of quantum error correction is to recover the encoded state
$\ket{\phi} \in \SSS$ from a possibly perturbed state $\ket{\phi'}
\in \HH$.  It turns out that it suffices to consider perturbations of
the form $\ket{\phi'} = \Eb\ket{\phi}$ for an \emph{error pattern}
$\Eb$ which is a Pauli $n$-tuple, because all possible linear
perturbations are linear combinations of Pauli $n$-tuples.

By the general principles of quantum mechanics, a measurement of an
operator of a quantum system yields an eigenvalue of that operator, and
projects the system state onto the corresponding
eigenspace.  A set of operators may be measured simultaneously if and
only if they commute.

In decoding, we first measure simultaneously the
$n-k$ commuting generators $\Gb_j$ of $S$.  This yields an $(n-k)$-tuple
of eigenvalues $\pm 1$, which may be mapped to a binary $(n-k)$-tuple
using the standard map $\{\pm 1\}  \to \{0, 1\}$.  The resulting binary
$(n-k)$-tuple $\sb = (s_1, \ldots, s_{n-k}) \in (\Z_2)^{n-k}$ is called
the \emph{syndrome sequence}.  This ``hard decision" in fact extracts
all relevant information.

If the error pattern is $\Eb$, then measurement of $\Gb_j$
results in the eigenvalue $+1$ if $\Eb$ commutes with $\Gb_j$, or $-1$
if $\Eb$ anticommutes with $\Gb_j$, regardless of the original state
$\ket{\phi} \in \SSS$.
(\emph{Proof}: If $\Eb$ commutes with
$\Gb_j$, then $\Gb_j(\Eb\ket{\phi}) = \Eb\Gb_j\ket{\phi} =
\Eb\ket{\phi}$;  otherwise
$\Gb_j(\Eb\ket{\phi}) = -\Eb\Gb_j\ket{\phi} = -\Eb\ket{\phi}$.)

Thus the syndrome bit $s_j$ depends only on $\Eb$, and
is given by $s_j(\Eb) = \Tr \inner{L(\Eb)}{L(\Gb_j)}$.
The syndrome bit map $s_j: [\Pi^n] \to \Z_2$ is a
homomorphism, by the bilinearity of the trace inner product.

The \emph{syndrome map} $\sb:  [\Pi^n] \to (\Z_2)^{n-k}$ defined by
$\sb(\Eb) = \{s_j(\Eb), 1 \le j \le n-k\}$ is thus a
homomorphism.  Its kernel is the set of all equivalence classes of Pauli
$n$-tuples that commute with all generators $\Gb_j$, which is precisely
the normalizer group $N(S)$.  It follows that each of the $2^{n-k}$
cosets of $N(S)$ in $[\Pi^n]$ maps to a
distinct binary syndrome $(n-k)$-tuple $\sb$.

In the case where $L(S)$ is $\F_4$-linear, we can compute an
$\F_4$-syndrome equal to the Hermitian inner product
$\inner{L(\Eb)}{\gb} \in \F_4$ for any $\gb \in L(S)$ as follows.
By $\F_4$-linearity, both $\omega \gb$ and
$\overline{\omega} \gb$ are also in $L(S)$.  The corresponding pair of
syndrome bits is
$$
(\Tr \inner{L(\Eb)}{\omega \gb}, \Tr \inner{L(\Eb)}{\overline{\omega}
\gb}) = (\Tr \overline{\omega}\inner{L(\Eb)}{\gb},  \Tr \omega
\inner{L(\Eb)}{\gb}).
$$
Now observe that $(\Tr \omega a, \Tr \overline{\omega} a) = \{(0,0), (1,
1), (1, 0), (0, 1)\}$ for $a = \{0, 1, \omega,
\overline{\omega}\}$;  in other words, $(\Tr \omega a, \Tr
\overline{\omega} a) =
\ell(a), a \in \F_4$.  Thus these two syndrome bits are the two bits of
the binary label
$\ell(\inner{L(\Eb)}{\gb})$, which identifies the
$\F_4$-syndrome  $\inner{L(\Eb)}{\gb}$.

Thus we can measure the $\F_4$-syndromes $S_k(\Eb) =
\inner{L(\Eb)}{\gb_k}$ for each of the $(n-k)/2$ generators $\gb_k$
of the quaternary stabilizer label code $L(S)$.  The syndrome map then
becomes a homomorphism
$\Sb:  [\Pi^n] \to (\F_4)^{(n-k)/2}$ with kernel $N(S)$.  Again, each of
the $4^{(n-k)/2}$ cosets of $N(S)$ in
$[\Pi^n]$ maps to a unique $\F_4$-syndrome $(n-k)/2$-tuple $\Sb$.

The syndrome measurement projects the perturbed state
$\ket{\phi'}$ onto a state $\ket{\phi''}$ in the
$2^k$-dimensional eigenspace $\SSS(\sb)$ that corresponds to the
measured syndrome $\sb$ (or $\Sb$).  If the actual perturbation was a
Pauli
$n$-tuple
$\Eb$, then $\ket{\phi'} = \Eb\ket{\phi}$ is an eigenvector of
$\SSS(\sb)$, so
$\ket{\phi''} = \ket{\phi'}$.

The next step in decoding is to find the most likely error pattern
$\hat{\Eb}$ in the coset of $N(S)$ whose syndrome is $\sb$ (or $\Sb$),
called the \emph{coset leader}.  The most likely error pattern is
assumed
to be the one of lowest Hamming weight;  \ie the Pauli $n$-tuple with
the fewest non-identity components.

Finding the coset leader is an entirely classical computation.  There
are $2^{n-k}$ possible syndromes $\sb$ (or $\Sb$), so if $n-k$ is not
too large, then the coset leaders may be precomputed, and this step may
be performed by a table lookup in a table with $2^{n-k}$ entries.   In
the QECC literature, the table lookup method is usually
assumed, explicitly or implicitly.

Finally, given the coset leader $\hat{\Eb}$, the perturbed state
$\ket{\phi'}$ is ``corrected" to $\hat{\Eb}\ket{\phi'} =
\hat{\Eb}\Eb\ket{\phi}$.  Decoding is successful if the coset leader
$\hat{\Eb}$ is merely in the same coset of
$S$ as the actual error pattern $\Eb$, so $\hat{\Eb}\Eb \in S$,
because any error pattern
$\hat{\Eb}\Eb \in S$ stabilizes any $\ket{\phi} \in \SSS$; \ie error 
patterns in
$S$ do not affect states in $\SSS$.
%In particular, if $\hat{\Eb}$ is in the equivalence class $[\Eb]$,
%then decoding is successful.

The set of $2^{n-k}$ coset leaders of the cosets of $N(S)$ (including
$[\Ib]$) will be denoted by
$\Sigma$.  The set of correctable error patterns is then $[S \Sigma]$.
If $[S \Sigma]$ contains all error patterns of Hamming weight $t$ or
less, then $\SSS$ is called a $t$-error-correcting code.

Since the Hamming distance function is a
true metric, $\SSS$ will be $t$-error-correcting if the minimum
Hamming distance $d$ between cosets of $S$ in $N(S)$ is greater than
$2t$ (\ie if $d \ge 2t+1$), because then two error patterns $\Eb$ and
$\hat{\Eb}$ of weight $\le t$ cannot lie in the same coset of
$N(S)$ unless they are in the same coset of $S$;  \ie $\hat{\Eb}\Eb$
cannot be in $N(S)$ unless it is in $S$.
     By the group property, this minimum
distance $d$ is the minimum Hamming weight of any nonzero coset of $S$
in $N(S)$;  \ie $d$ is the minimum Hamming weight in $N(S)\setminus S$.

An $[n,k]$ stabilizer code in which $N(S)\setminus S$
has minimum Hamming weight $d$ is called an $[n,k,d]$ stabilizer
code.  We will consider only \emph{nondegenerate} codes, in which the
normalizer code
$L(N(S))$ actually has minimum Hamming distance $d$;  \ie $L(S)$ has
a minimum distance of at least $d$.

\subsection{Summary:  stabilizer codes from $\F_4$-linear codes}

In summary, to construct a nondegenerate $[n,k,d]$ stabilizer code with
$n-k$ even, it suffices to find a classical self-orthogonal $(n,
(n-k)/2)$ $\F_4$-linear block code $L(S)$ whose orthogonal
$(n, (n+k)/2)$ code
$L(S)^\perp$ under the Hermitian inner product has minimum Hamming
distance $d^\perp = d$.

\smallskip
\noindent
\textbf{Example A} (Five-qubit ``quantum Hamming code").
In order to construct a single-error-correcting $[5,1,3]$ stabilizer
code, we take the quaternary stabilizer label code $L(S)$ to be the
classical $(5, 2, 4)$ self-orthogonal (doubly extended Reed-Solomon)
code
over $\F_4$ generated by
$$
\begin{array}{ccccc}
0 & \overline{\omega} & \omega & \omega & \overline{\omega} \\
\overline{\omega} & 0 & \overline{\omega} & \omega & \omega
\end{array},
$$
The orthogonal code $L(N(S)) = L(S)^\perp$ under the
Hermitian inner product is then the  $(5, 3, 3)$ quaternary Hamming
code.

A $[5,1,3]$ stabilizer code is single-error-correcting.
Each of the 15 error patterns in $\Pi^5$ of Hamming weight 1 therefore
lies in a distinct one of the 15 nonzero cosets of $N(S)$;  \ie this
code is a ``perfect" single-error-correcting ``quantum Hamming
code."  Decoding may be performed by a table lookup in a table that maps
each of the 16 possible syndromes to the corresponding minimum-weight
error pattern.
\qed

Similarly, for all integers $m \ge 2$, there exist classical perfect
$\F_4$-linear Hamming codes with parameters $(n = (4^m - 1)/3, k = n -
m, d = 3)$ that contain their orthogonal $(n, m)$ codes
\cite{CRSS98}.\footnote{By Bonisoli's theorem \cite{W92}, the  $(n,m)$
code is equidistant, with all nonzero codewords having weight
$4^{m-1}$.}  These codes may be used to construct quantum $[n = (4^m -
1)/3, k = n - 2m, d = 3]$ Hamming codes that can be decoded using a
lookup table with
$4^m$ entries;  \eg stabilizer codes with parameters $[5, 1, 3], [21,
15, 3], [85, 77, 3]$, and so forth.

\subsection{CSS-type stabilizer codes from $\F_2$-linear codes}

The binary field $\F_2$ is a subfield of the quaternary field $\F_4$.
Therefore the $(n-k)/2$ generators $\{\gb_j, 1 \le j \le (n-k)/2\}$ of a
classical $(n, (n-k)/2)$ binary linear code $\B$ may be taken as the
generators of an $(n, (n-k)/2)$ $\F_4$-linear code $\CC$.  Since the
Hermitian inner product of binary sequences is the ordinary binary inner
product, $\CC$ will be self-orthogonal if $\B$ is self-orthogonal.

As we have seen, $\CC$ may be characterized as the set of all binary
linear combinations of the $n-k$ generators $\{\omega
\gb_j,\overline{\omega} \gb_j \mid 1 \le j \le (n-k)/2\}$.  The two-bit
labels $\ell(\omega \gb_j)$ are nonzero only in bit flip
bits, whereas the labels $\ell(\overline{\omega} \gb_j)$ are
nonzero only in phase flip bits.  Therefore the binary code
$\ell(\CC)$ may be characterized as two interleaved, independent binary
codes:  namely, the code $\B$ applied to the $n$ bit flip bits, and the
code $\B$ applied to the $n$ phase flip bits, respectively.  In short,
$\ell(\CC)$ is a direct product code:
$$
\ell(\CC) = \B \times \B.
$$

Similarly, the orthogonal code $\CC^\perp$ to $\CC$ is generated by
$(n+k)/2$ generators of the orthogonal $(n, (n+k)/2)$ binary linear code
$\B^\perp$, and the corresponding binary code is the direct product
code $\ell(\CC^\perp) = \B^\perp \times \B^\perp$.
If the minimum distance of
$\B^\perp$ is $d^\perp$, then the minimum distance of $\CC^\perp$ is
$d^\perp$.

More generally, Calderbank and Shor \cite{CS96} and Steane \cite{S96}
proposed codes of the form  $\B_1 \times
\B_2$, where the bit flip code $\B_1$ and the phase flip code $\B_2$ are
possibly different orthogonal binary codes.
%Such codes are not in general $\F_4$-linear, if $\B_1 \neq \B_2$.
We will consider only  codes of the type $\B \times \B$,
which we will call \emph{CSS-type codes}.

In short, to construct a nondegenerate $[n,k,d]$ stabilizer code with
$n-k$ even, it suffices to find a classical self-orthogonal $(n,
(n-k)/2)$ binary linear block code $\B$ whose orthogonal
$(n, (n+k)/2)$ code $\B^\perp$ has minimum Hamming
distance $d^\perp = d$.

\smallskip
\noindent
\textbf{Example B} (Seven-qubit Steane code)
Consider the $(7,3,4)$ binary linear (dual Hamming) code $\B$ that is
generated by the following generators:
$$
\begin{array}{ccccccc}
0 & 0 & 0 & 1 & 1 & 1 & 1 \\
0 & 1 & 1 & 0 & 0 & 1 & 1 \\
1 & 0 & 1 & 0 & 1 & 0 & 1
\end{array}.
$$
This code is evidently self-orthogonal.  Its orthogonal code $\B^\perp$
is the $(7,4,3)$ binary Hamming code. Thus the resulting
stabilizer code is a $[7,1,3]$ single-error-correcting code.  \qed

In general, CSS-type codes have poorer parameters
$[n,k,d]$ than general $\F_4$-linear codes, because binary codes have
poorer parameters than quaternary codes.  However, they have the
advantage that bit flip and phase flip errors may be decoded separately,
as follows.  Note that the syndrome bits may be written as
$$
(\Tr \inner{L(\Eb)}{\omega \gb_j}, \Tr \inner{L(\Eb)}{\overline{\omega}
\gb_j}) = (\inner{\ell_2(\Eb)}{\gb_j}, \inner{\ell_1(\Eb)}{\gb_j}),
$$
because $\ell(\omega \gb_j) = (\gb_j, \zerob)$ and
$\ell(\overline{\omega} \gb_j) = (\zerob, \gb_j)$.  In other words, the
first syndrome bits in each pair form a set of $(n-k)/2$ syndromes for
the phase error bits, and the second for the bit error bits.  We can
then decode each set of syndromes independently, using a decoder for the
binary code $\B^\perp$.

If the Hamming weight of $\Eb$ is not
greater than $\lfloor (d-1)/2 \rfloor$, then the Hamming weights of
$\ell_1(\Eb)$ and $\ell_2(\Eb)$ both satisfy the same bound, so decoding
will be successful.  Indeed, two independent binary decodings of up to
$\lfloor (d-1)/2 \rfloor$ bit errors will correct some higher-weight
error patterns.

\smallskip
\noindent
\textbf{Example B} (cont.) A quaternary decoder for the $(7,4,3)$
$\F_4$-linear code $\CC^\perp$ requires calculation of three
$\F_4$-syndromes and table lookup in a $64$-entry table (for complete
decoding;  single-error-correction requires at least 21 entries).  A
binary decoder for the $(7,4,3)$  binary linear code $\B^\perp$ requires
calculation of three  binary syndromes and table lookup in an $8$-entry
table.  Two binary decodings of
$\B^\perp$ are thus arguably simpler than one quaternary decoding of
$\CC^\perp$.  \qed

Similarly, for all integers $m \ge 3$, there exist classical perfect
binary Hamming codes with parameters $(n = 2^m - 1, k = n - m, d =
3)$ that contain their orthogonal $(n, m)$ codes.\footnote{By
Bonisoli's theorem, the $(n,m)$ code is equidistant, with
all nonzero codewords having weight $2^{m-1}$.}  These codes may be used
to
construct single-error-correcting quantum $[n = 2^m - 1, k = n - 2m, d =
3]$ stabilizer codes that can be decoded by using a lookup table with
$2^m$ entries twice;  \eg stabilizer codes with parameters
$[7, 1, 3], [15, 7, 3], [31, 21, 3]$, and so forth.

In summary, while CSS-type codes based on binary codes may have poorer
parameters $[n,k,d]$ than codes based on $\F_4$-linear codes, they may
nevertheless have advantages in terms of decoding complexity.
(The question of decoding  complexity seems hardly to have been
addressed previously in the QECC literature, with a few notable
exceptions,
\eg
\cite{BG98, GB00, MMM04}.) For example, the $[15, 7, 3]$ CSS-type
code has comparable decoding complexity to the $[5, 1, 3]$ $5$-qubit
quantum Hamming code, as well as comparable performance, but  has a
higher code rate.

\section{Quantum convolutional and tail-biting codes}

In this section, using the stabilizer formalism, we show how to
construct quantum convolutional and tail-biting codes from classical
self-orthogonal
$\F_4$-linear and binary convolutional codes.

\subsection{Infinite-qubit systems}
A quantum convolutional code will be defined (at least in principle) on a
quantum system containing a countably infinite ordered sequence of qubits; 
\ie whose Hilbert space $\HH =
\cdots\otimes\HH_{-1}\otimes\HH_0\otimes\HH_1\otimes\cdots$ is the tensor
product of an infinite sequence of two-dimensional Hilbert spaces
$\HH_i, i \in \Z$, and thus is infinite-dimensional.

 A Pauli sequence $\Ab=\bigotimes_{i\in\Z} A_i$ acting on $\HH$ is an
infinite tensor product of  Pauli matrices $A_i, i \in \Z$.  The set of all
such infinite Pauli sequences may be denoted by $\Pi^\Z$.
% A Pauli
%matrix $A$ acting only on the $i$th qubit (and trivial on the other
%qubits) will be denoted by $\Ab^{(i)}$.  
Again we may define the set $[\Pi^\Z] = \{\beta\Ab|\Ab \in \Pi^\Z, \beta \in
\C, |\beta| = 1\}$ of equivalence classes of Pauli sequences up to phase.
The label maps
$\ell$ and $L$ of Section~II.A may be extended to maps from $[\Pi^\Z]$
to sequences in $((\Z_2)^2)^\Z$ and $(\F_4)^\Z$, respectively.  Inner
products and convolutions are well defined, provided that the corresponding
sums involve only a finite number of nonzero terms.

\subsection{Convolutional stabilizer codes}

We will define a rate-$k/n$ convolutional stabilizer code, $0 \le k \le
n$, by a basic set of $n-k$ independent commuting Pauli sequences $\G_0 =
\{\Gb_j, 1 \le j \le n-k\}$, where each generator $\Gb_j \in \G$ has finite
support;  \ie only finitely many Pauli matrices in $\Gb_j$ are not 
identity matrices.  

The full generator set $\G$ will then be the set
of all shifts of the $k$ basic generators in $\G_0$ by integer multiples of
$n$ qubits.  We require the generators in $\G$ to be independent and
commuting.  The code subspace $\SSS$ will again be defined to be the
subspace of
$\HH$ that is stabilized by $\G$.

The stabilizer group $S$ is again defined as the subgroup of $[\Pi^\Z]$ of
all products of generators in $\G$.  The binary and quaternary label codes
$\ell(S)$ and $L(S)$ again denote the images of $S$ under the binary and
quaternary label maps, respectively.  

In this paper we will only consider codes in which
$L(S)$ is $\F_4$-linear.  Then $n-k$ must be even, and $L(S)$ is a
classical rate-$((n-k)/2)/n$ convolutional code over $\F_4$ generated by
the set of all shifts by integer multiples of $n$ symbols of some set of
$(n-k)/2$ independent generators $\{\gb_i, 1 \le i \le (n-k)/2\}$ with
finite support.  In this case the generators in $\G_0$ may be taken as the
inverse images of the $n-k$ generators $\omega\gb_i$ and $\omegabar
\gb_i$.  

Finally, in order that each generator in $\G_0$ commute with all shifts of
all generators in $\G_0$, the $\F_4$-linear convolutional code $L(S)$ must
be self-orthogonal under the Hermitian inner product;  \ie a subcode of its
orthogonal code $L(S)^\perp$, a classical $\F_4$-linear
rate-$((n+k)/2)/n$ convolutional code.

As with block stabilizer codes, a convolutional
stabilizer code will turn out to have minimum distance $d$ if the orthogonal
code $L(S)^\perp$ has minimum Hamming distance $d^\perp = d$.

We will focus on rate-$(n-2)/n$ convolutional stabilizer codes, for which
the rates of $L(S)$ and $L(S)^\perp$ are $1/n$ and $(n-1)/n$, respectively.
The generators of a rate-$1/n$ classical convolutional code $L(S)$ are the
set of all shifts by an integer number of $(\F_4)^n$-blocks of a single
finite-support sequence $\gb = \{g_{m} \in (\F_4)^n, m \in
\Z\}$.

\smallskip\noindent
\textbf{Example 1} (rate-1/3, single-error-correcting, $\F_4$-linear
convolutional stabilizer code).
Consider the classical rate-1/3 $\F_4$-linear shift-invariant
convolutional code $\CC = L(S)$ that is generated by all shifts by integer
multiples of 3 symbols of the generator sequence $\gb_1 = (\ldots| 000| 111|
1
\omega
\overline{\omega}| 000|
\ldots)$;  \ie whose generators are:
$$
\begin{array}{c}
\ldots \\
\begin{array}{c|ccc|ccc|ccc|ccc|ccc|ccc|c}
\ldots & 0 & 0 & 0 & 1 & 1 & 1 & 1 & \omega & \overline{\omega} & 0 & 0
&
0 & 0 & 0 & 0 & 0 & 0 & 0 & \ldots \\
\ldots & 0 & 0 & 0 & 0 & 0 & 0 & 1 & 1 & 1 & 1 & \omega &
\overline{\omega} & 0 & 0 & 0 & 0 & 0 & 0 & \ldots \\
\ldots & 0 & 0 & 0 & 0 & 0 & 0 & 0 & 0 & 0 & 1 & 1 & 1 & 1 & \omega &
\overline{\omega} & 0 & 0 & 0 & \ldots \\
\end{array} \\
\ldots
\end{array}
$$
In $D$-transform notation (see Section IV), the basic generator is
$\gb_1(D) = (1 + D, 1 + \omega D, 1 + \overline{\omega} D)$, and the set
of all generators is
$\{D^\ell \gb_1(D), \ell \in \Z\}$.

    The corresponding stabilizer group
$S$ is then generated by sequences of Pauli matrices that correspond
to multiples by $\omega$ and $\overline{\omega}$ of these
generators;  \ie the generators of $S$ are the shifts by an integral
number of $3$-blocks of the two basic generators $(\ldots| III| XXX| 
XZY|
III| \ldots)$ and  $(\ldots| III| ZZZ| ZYX| III| \ldots)$.

It is easy to verify that $\gb_1$ is orthogonal to itself and to any
shift of itself under the Hermitian inner product, which suffices to
show that all generators are orthogonal.  Thus $\CC$ is
self-orthogonal;  \ie all generators of $S$ commute.

The orthogonal convolutional code $\CC^\perp$ under the
Hermitian inner product is the rate-2/3  $\F_4$-linear
convolutional code that is generated by all shifts of $\gb_1$ and $\gb_2
= (\ldots| 000| \overline{\omega} \omega 1| 000| \ldots)$, whose
$D$-transform is $\gb_2(D) = (\overline{\omega}, \omega, 1)$. It is easy
to verify that the minimum Hamming distance of
$\CC^\perp$ is $d^\perp = 3$, with the only weight-3 codewords being
multiples of shifts of $\gb_2$. The convolutional stabilizer code
defined
by $\CC$ thus has minimum Hamming distance 3, so it is
single-error-correcting.
\qed

In principle, the convolutional code $\CC$ of Example 1 has an
infinite number of generators $\gb_j$, covering an infinite
number of $3$-blocks.   However, because the support of
each generator is only two $3$-blocks, the code constraints are
localized;  the code symbols in any block depend only on the ``current"
and ``previous" generators.  Such a convolutional code is said to have a
``memory" or \emph{constraint length} of one $3$-block ($\nu = 1$).

\subsection{Block codes from convolutional codes}

In data communications, where information symbols are often transmitted
in a continuous stream, the non-block structure of convolutional codes
is often a virtue rather than a problem.  However, for the main
applications currently envisioned for quantum error-correcting codes,
such as protection of the state of a quantum computer, a block structure
is desirable.  In this subsection, we will discuss two methods of making
a convolutional code into a block code:  termination and tail-biting.

To construct a terminated block code $\B$ from a convolutional
code $\CC$, we simply take the generators of $\B$ to be the subset of
all generators of $\CC$ whose support lies in some given interval.
Since $\B$ is a subcode of $\CC$, it will be self-orthogonal if $\CC$ is
self-orthogonal, and its minimum distance must be at least as great as that
of $\CC$. The rate of $\B$ will in general be less than that of $\CC$, but
it will approach the rate of $\CC$ as the length of the interval increases.  

The orthogonal block code $\B^\perp$ will then be the code generated by the
truncations of the generators of the orthogonal convolutional code
$\CC^\perp$ to the same interval.  $\B^\perp$ may therefore have a smaller
minimum distance than
$\CC^\perp$, and may not be self-orthogonal even if $\CC^\perp$
is self-orthogonal.

There are therefore two ways that we might think of terminating a
convolutional stabilizer code defined by a classical $\F_4$-linear
convolutional code $\CC = L(S)$.  We might try terminating $\CC^\perp$ to a
block code $\B^\perp$, thus keeping the minimum distance of $\B^\perp$
equal to that of
$\CC^\perp$.  However, the dual block code $\B$ will then be a truncation of
the convolutional code $\CC$, which is not guaranteed to be
self-orthogonal, so in general $\B$ will not define a valid block
stabilizer code.  The alternative is to terminate $\CC$ to a block code
$\B$, thus ensuring self-orthogonality, but in this case the truncated code
$\B^\perp$ is not guaranteed to have the same minimum distance as
$\CC^\perp$.

\smallskip\noindent
\textbf{Example 1} (cont.) For example, taking the convolutional code as
the rate-1/3 convolutional code $\CC$ of Example 1 and an interval
of  three consecutive $3$-blocks, the following two generators of $\CC$
have support contained in the given interval:
$$
\begin{array}{ccc|ccc|ccc}
%\overline{\omega} & \omega & 1 & 0 & 0 & 0 & 0 & 0 & 0 \\
1 & 1 & 1 & 1 & \omega & \overline{\omega} & 0 & 0 & 0 \\
%0 & 0 & 0 & \overline{\omega} & \omega & 1 & 0 & 0 & 0 \\
0 & 0 & 0 & 1 & 1 & 1 & 1 & \omega & \overline{\omega} \\
%0 & 0 & 0 & 0 & 0 & 0 & \overline{\omega} & \omega & 1
\end{array}
$$
These generate a $(9, 2)$ $\F_4$-linear terminated block code
$\B$, which is evidently self-orthogonal.

The orthogonal block code $\B^\perp$ is the $(9, 7)$ block
code generated by the following seven truncated generators of
$\CC^\perp$:
$$
\begin{array}{ccc|ccc|ccc}
1 & \omega & \overline{\omega} & 0 & 0 & 0 & 0 & 0 & 0 \\
\overline{\omega} & \omega & 1 & 0 & 0 & 0 & 0 & 0 & 0 \\
1 & 1 & 1 & 1 & \omega & \overline{\omega} & 0 & 0 & 0 \\
0 & 0 & 0 & \overline{\omega} & \omega & 1 & 0 & 0 & 0 \\
0 & 0 & 0 & 1 & 1 & 1 & 1 & \omega & \overline{\omega} \\
0 & 0 & 0 & 0 & 0 & 0 & \overline{\omega} & \omega & 1 \\
0 & 0 & 0 & 0 & 0 & 0 & 1 & 1 & 1 
\end{array}
$$
The minimum distance of $\B^\perp$ is now only 2 (\eg the sum of the
first two generators has weight 2).

Alternatively, taking the convolutional code as
the rate-2/3 convolutional code $\CC^\perp$ of Example 1 and the same
three-block interval, the following five generators of
$\CC^\perp$ have support contained in the given interval:
$$
\begin{array}{ccc|ccc|ccc}
\overline{\omega} & \omega & 1 & 0 & 0 & 0 & 0 & 0 & 0 \\
1 & 1 & 1 & 1 & \omega & \overline{\omega} & 0 & 0 & 0 \\
0 & 0 & 0 & \overline{\omega} & \omega & 1 & 0 & 0 & 0 \\
0 & 0 & 0 & 1 & 1 & 1 & 1 & \omega & \overline{\omega} \\
0 & 0 & 0 & 0 & 0 & 0 & \overline{\omega} & \omega & 1
\end{array}
$$
These generate a $(9, 5, 3)$ $\F_4$-linear terminated block code
$\B^\perp$.

The orthogonal block code is the $(9, 4)$ $\F_4$-linear block code $\B$
generated by the following four truncated generators of
$\CC$:
$$
\begin{array}{ccc|ccc|ccc}
1 & \omega & \overline{\omega} & 0 & 0 & 0 & 0 & 0 & 0 \\
1 & 1 & 1 & 1 & \omega & \overline{\omega} & 0 & 0 & 0 \\
0 & 0 & 0 & 1 & 1 & 1 & 1 & \omega & \overline{\omega} \\
0 & 0 & 0 & 0 & 0 & 0 & 1 & 1 & 1
\end{array}
$$
Now $\B$ is not even self-orthogonal, because the truncated
generators are not self-orthogonal.  \qed

For our purposes, \emph{tail-biting} (see, \eg \cite{CFV99}) is a better
method of making a convolutional code into a block code.
To construct a tail-biting block code $\B$ from a
convolutional code $\CC$, we take the generators of $\B$ to be
the subset of all generators of $\CC$ whose ``starting time" lies
in some given interval.  If the ``ending time" lies outside
the given interval, then we wrap the generator around to the
beginning of the interval in ``tail-biting" fashion;  see the example
below.   We assume that the length of the tail-biting interval is
greater than that of the support of any generator.

It can be easily shown that the rate of $\B$ will be the same as
   that of $\CC$ if the generators are noncatastrophic (see Section IV).
There is now no guarantee that the minimum distance of $\B$ will be as
great as that of $\CC$; however, in general the minimum distance will
be preserved if the tail-biting interval is long enough
\cite{HHJZ02}. The orthogonal block code $\B^\perp$ will be the
corresponding  tail-biting block code derived from $\CC^\perp$.
Finally, if $\CC$ is self-orthogonal, then $\B$ will be self-orthogonal.

A tail-biting code $\B$ derived from a self-orthogonal convolutional
code $\CC$ may therefore be used to specify a block stabilizer code
with the same rate as the convolutional stabilizer code derived from
$\CC$, and, provided that the block length is large enough, the same
minimum distance.

\smallskip\noindent
\textbf{Example 2} (rate-$1/3$, single-error-correcting, $\F_4$-linear
tail-biting stabilizer code).  If we again take the convolutional code as
the rate-$1/3$ convolutional code
$\CC$ of Example 1 and a tail-biting interval of length three
3-blocks, then we obtain the following three tail-biting generators:
$$
\begin{array}{ccc|ccc|ccc}
1 & 1 & 1 & 1 & \omega & \overline{\omega} & 0 & 0 & 0 \\
0 & 0 & 0 & 1 & 1 & 1 & 1 & \omega & \overline{\omega} \\
1 & \omega & \overline{\omega} & 0 & 0 & 0 & 1 & 1 & 1 \\
\end{array}
$$
Notice how the last generator has been ``wrapped around."
Thus these generators generate a $(9, 3)$ $\F_4$-linear tail-biting code
$\B$.
Moreover, since $\CC$ is self-orthogonal, $\B$ is self-orthogonal.

\pagebreak
The orthogonal block code is the $(9, 6)$ $\F_4$-linear tail-biting
code $\B^\perp$ that is generated by the following six tail-biting
generators:
$$
\begin{array}{ccc|ccc|ccc}
\overline{\omega} & \omega & 1 & 0 & 0 & 0 & 0 & 0 & 0 \\
1 & 1 & 1 & 1 & \omega & \overline{\omega} & 0 & 0 & 0 \\
0 & 0 & 0 & \overline{\omega} & \omega & 1 & 0 & 0 & 0 \\
0 & 0 & 0 & 1 & 1 & 1 & 1 & \omega & \overline{\omega} \\
0 & 0 & 0 & 0 & 0 & 0 & \overline{\omega} & \omega & 1 \\
1 & \omega & \overline{\omega} & 0 & 0 & 0 & 1 & 1 & 1 \\
\end{array}
$$
The minimum distance of $\B^\perp$ turns out to be $d^\perp = 3$, so
$\B$ defines a $[9, 3, 3]$ block stabilizer code.
\qed

\subsection{Decoding algorithms}

We now discuss how to decode a convolutional stabilizer code
that has been constructed from a classical self-orthogonal $\F_4$-linear
rate-$k/n$ convolutional code $\CC$.

As shown in Section II, we may first measure each
generator $\gb_j$ of the convolutional code $\CC$ to obtain a sequence
$\Sb$ of
$\F_4$-syndromes $S_j = \inner{L(\Eb)}{\gb_j} \in \F_4$, where $L(\Eb)$
denotes the quaternary error label sequence $L(\Eb)$, at a rate of $k$
$\F_4$-syndromes for each $n$-block.  The syndrome sequence $\Sb$
determines a coset $\CC^\perp + \tb(\Sb)$ of the orthogonal
convolutional
code $\CC^\perp$, where $\tb(\Sb)$ is any error sequence whose syndrome
sequence is $\Sb$.\footnote{For example, if $\Sb = \eb H^T$ and
$(H^{-1})^T$ is any left inverse of $H^T$, then we may take $\tb(\Sb) =
\Sb (H^{-1})^T$.}  We then need to find the minimum-weight coset leader
in that coset, which is an entirely classical computation.

A standard way of finding the leader of a coset $\CC^\perp + \tb(\Sb)$
of a convolutional code $\CC^\perp$ is to represent the coset by a trellis
diagram in which there is one-to-one correspondence between
coset sequences and trellis paths, and then search for the
lowest-weight trellis path by the Viterbi algorithm (VA) \cite{F73b}.
The trellis diagram may be taken as any trellis diagram for $\CC^\perp$,
with all code sequences translated by the representative error sequence
$\tb(\Sb)$.

For example, the rate-$2/3$ convolutional code $\CC^\perp$ of Example 1
has a minimal trellis diagram with 4 states at each $3$-block boundary,
and  64 transitions between trellis states during each $3$-block.  A
VA search through this trellis requires of the
order of 64 computations  per 3-block.

If our objective is merely correction of single errors, however, then we
can use the following much simpler algorithm.  As long as
all syndromes are zero, we assume that no errors have occurred.  Then,
if a nonzero syndrome $S_j$ occurs, we assume that a single error has
occurred in one of the three qubits in the $j$th block;  the error
is characterized by a label $3$-tuple $\eb_j = L(\Eb_j)$.  The nine
possible weight-1 error $3$-tuples
$\eb_j$ lead to the following syndromes
$(S_j,S_{j+1})$ during blocks $j$ and $j+1$:

$$
\begin{array}{c|c}
\eb_j & (S_j, S_{j+1}) \\
\hline
1 0 0  & (1, 1) \\
\omega 0 0 & (\omega, \omega) \\
\overline{\omega} 0 0 & (\overline{\omega}, \overline{\omega}) \\
0 1 0  & (\overline{\omega}, 1) \\
0 \omega 0 & (1, \omega) \\
0 \overline{\omega} 0 & (\omega, \overline{\omega}) \\
0 0 1  & (\omega, 1) \\
0 0 \omega & (\overline{\omega}, \omega) \\
0 0 \overline{\omega} & (1, \overline{\omega})
\end{array}
$$
Since these 9 syndrome pairs are distinct, we may map
$(S_j,S_{j+1})$ to the corresponding single-error label $3$-tuple
$\eb_j$ using a simple $9$-entry table lookup, and then correct the
error as indicated.  (If $(S_j,S_{j+1})$ is not in the table---
\ie if $S_{j+1} = 0$--- then we have detected a weight-2 error.)

We see that this simple algorithm can correct any single-error
pattern $\Eb_j$, provided that there is no second error during blocks
$j$ and $j+1$.   The decoder synchronizes itself properly whenever
a zero syndrome occurs, and subsequently can correct one error in every
second block, provided that every errored block is followed by an
error-free block.

To decode the $(9,6,3)$ tail-biting code $\B^\perp$ of Example 2, we may
use the same algorithm, but now on a ``circular" time axis.
Specifically, if only a single error occurs, then one of the three
resulting $\F_4$-syndromes will be zero, and the other two nonzero.  The
zero syndrome tells which block the error is in;  the remaining two
nonzero syndromes determine the error pattern according to the $9$-entry
table given above.  Thus again we need only a $9$-entry table lookup.
(Notice that the existence of a single-error-correcting decoder for
$\B^\perp$ proves that its minimum distance is $d^\perp = 3$.)

More generally, there are several methods of adapting the Viterbi
algorithm to decode tail-biting codes \cite{CFV99}.  The tail-biting
code
trellis diagram may be taken as a finite-length segment of the
corresponding convolutional code trellis, with the further constraint
that valid paths must start and end in the same state.  An optimal
decoding method is VA decoding of each subtrellis consisting of all
paths starting and ending in a given state, followed by selection of the
best of these decoded paths.  A simpler suboptimal method is to regard
the tail-biting trellis as being defined on a circular time axis, and to
use a single VA search from an arbitrary initial state around and
around the circular trellis, until convergence is obtained.

\subsection{Comparison of single-error-correcting codes}

We now briefly compare the rate, performance and decoding
complexity of our single-error-correcting convolutional and
tail-biting codes with those of comparable previous unrestricted QECCs.

First, we will compare the decoding error probability per encoded qubit
of our convolutional and tail-biting codes to
that of the single-error-correcting $5$-qubit block code (Example A).  
We
assume that the probability of an error in any qubit is
$p$, independent of errors in other qubits.  Our estimates do not depend
on the relative probabilities of $X, Y$ or $Z$ errors.

For the $5$-qubit block code, a decoding error occurs if
there are 2 errors in any block, so the error probability is of
the order of ${5 \choose 2} p^2 = 10p^2$ per block, or per encoded
qubit.

For our rate-$1/3$ convolutional code, for each $3$-qubit block,
a decoding error may occur if there are 2 errors in that block, or 1 in
that block and 1 in the subsequent block.  The error probability is
therefore of the order of $(3 + 3^2) p^2 = 12 p^2$ per $3$-qubit block, 
or per encoded qubit.

Finally, for our $[9, 3, 3]$ tail-biting code, a decoding error
may occur if there are 2 errors in a block of 9 qubits, so the error
probability is of the  order of ${9 \choose 2} p^2 = 36p^2$ per block,
or $12p^2$ per encoded qubit.

We conclude that the decoding error probability per encoded qubit is
very nearly the same for any of these three codes.

With regard to rate and decoding complexity, our quantum convolutional
and tail-biting codes have rate $1/3$, which is greater than that of any
previous simple single-error-correcting quantum code, block or
convolutional.  Our decoding algorithm involves only a $9$-entry table
lookup, which is at least as simple as that of any previous quantum
code.

Our convolutional code rate and error-correction capability (one error
every two $3$-blocks) are comparable to those of a $[6, 2, 3]$ block
stabilizer code.  However, no $[6, 2, 3]$ block stabilizer code exists
(by the ``quantum Hamming bound," since it  would be a nondegenerate
quantum MDS code).

Our tail-biting code is a $[9,3,3]$ block stabilizer code.  We could
obtain a code with the same parameters by shortening a
$[21, 15, 3]$ quantum Hamming code.  However, such a shortened code
would not necessarily have such a simple structure as our
tail-biting code, nor such a simple decoding algorithm.

\subsection{CSS-type convolutional codes}

Similarly, as with CSS-type block codes, we may construct a
CSS-type convolutional stabilizer code with minimum distance $d$ from a
classical self-orthogonal binary convolutional code $\CC$ whose
orthogonal convolutional code $\CC^\perp$ has
minimum Hamming distance $d^\perp = d$.
Again, if the rate of $\CC$ is $(n-k)/2n$, then the
rate of $\CC^\perp$ will be $(n+k)/2n$, and the rate of the
convolutional stabilizer code will be $k/n$.  We continue to focus on
rate-$1/n$ codes.

As with CSS-type block codes, we will find that the parameters of
CSS-type convolutional codes are in general poorer than those of codes
based on general $\F_4$-linear codes, but that they may offer complexity
advantages, since bit flip errors and phase flip errors may be decoded
by two independent decodings of the binary convolutional code $\CC^\perp$.

\smallskip\noindent
\textbf{Example 3} (rate-$1/3$, single-error-correcting, CSS-type
convolutional code).
Consider the binary rate-$1/3$
convolutional code $\CC$ whose generators are the shifts by an integral
number of $3$-blocks of the single basic generator
$\gb = (\ldots| 000| 111| 100| 110| 000| \ldots)$, whose $D$-transform
is $\gb(D) = (1 + D + D^2, 1 + D^2, 1)$:

$$
\begin{array}{c}
\ldots \\
\begin{array}{c|ccc|ccc|ccc|ccc|ccc|ccc|c}
\ldots & 0 & 0 & 0 & 1 & 1 & 1 & 1 & 0 & 0 & 1 & 1 & 0 & 0 & 0 & 0 & 0 &
0 & 0 & \ldots \\
\ldots & 0 & 0 & 0 & 0 & 0 & 0 & 1 & 1 & 1 & 1 & 0 & 0 & 1 & 1 & 0 & 0 &
0 & 0 & \ldots
\end{array} \\
\ldots
\end{array}
$$
The ``memory" of $\CC$ is thus two $3$-blocks (\ie its constraint length
is $\nu = 2$).

    The  stabilizer group $S$ is then generated by sequences of
Pauli matrices that correspond to multiples of the above generators by
$\omega$ and $\overline{\omega}$.  Thus the generators of $S$ are the
shifts by an integral number of $3$-blocks of two basic generators,
$(\ldots| III| XXX| XII| XXI| III| \ldots)$ and  $(\ldots| III| ZZZ|
ZII| ZZI| III| \ldots)$.  Since these stabilizers affect only bit
flip and phase flip bits, respectively, the code $\CC$ is the direct
product of two independent binary codes that protect the bit flip and
phase flip bits, respectively.

It is easy to verify that $\gb$ is orthogonal to itself and to any
shift of itself under the usual binary inner product.  This suffices
to show that $\CC$ is self-orthogonal.

The generators of the orthogonal rate-$2/3$ binary convolutional code
$\CC^\perp$ are the shifts of two basic generators, $\hb_1 = (\ldots|
000| 110| 011| 000| \ldots)$ and $\hb_2 = (\ldots| 000| 001| 110| 000|
\ldots)$, whose $D$-transforms are $\hb_1(D) = (1, 1 + D, D)$ and
$\hb_2(D) = (D, D, 1)$, respectively:
$$
\begin{array}{c}
\ldots \\
\begin{array}{c|ccc|ccc|ccc|ccc|ccc|c}
\ldots & 0 & 0 & 0 & 1 & 1 & 0 & 0 & 1 & 1 & 0 & 0 & 0 & 0 & 0 & 0 &
\ldots \\
\ldots & 0 & 0 & 0 & 0 & 0 & 1 & 1 & 1 & 0 & 0 & 0 & 0 & 0 & 0 & 0 &
\ldots \\
\ldots & 0 & 0 & 0 & 0 & 0 & 0 & 1 & 1 & 0 & 0 & 1 & 1 & 0 & 0 & 0 &
\ldots \\
\ldots & 0 & 0 & 0 & 0 & 0 & 0 & 0 & 0 & 1 & 1 & 1 & 0 & 0 & 0 & 0 &
\ldots \\
\end{array} \\
\ldots
\end{array}
$$
It is easy
to verify that the minimum Hamming distance of
$\CC^\perp$ is $d^\perp = 3$, with the only weight-3 codewords being
shifts of $\hb_2$. Thus the rate-$1/3$ convolutional stabilizer code
defined by $\CC$ has minimum Hamming distance 3, and is
single-error-correcting.  \qed

We now consider how to decode the Example 3 code.  We will discuss only
how to decode bit flip errors;  phase flip errors may be corrected
independently and identically.

For bit flip errors, we first measure each generator
$\gb_j$ of $\CC$ to obtain a sequence $\sb$ of  binary syndromes $s_{j}
= \inner{\ell_1(\Eb)}{\gb_j} \in \F_2$, where
$\inner{\ell_1(\Eb)}{\gb_j}$ denotes the binary inner product of
the generator
$\gb_j$ with the bit flip error label sequence $\ell_1(\Eb)$, at
a rate of one binary syndrome for each $3$-block.

Again, instead of VA decoding the 4-state trellis of the rate-$2/3$ code
$\CC^\perp$, we may use a simple single-error-correction algorithm, as
follows.  As long as all syndromes are zero, we assume that no errors
have occurred.  When a nonzero syndrome $s_{j}$ occurs, we assume
that a single error has occurred in one of the three bit flip bits in
block $j$, corresponding to a binary label $3$-tuple $\eb_j =
\ell_1(\Eb_j)$.  The three possible weight-1 error $3$-tuples
$\eb_j$ lead to the following syndrome sequences:
$$
\begin{array}{c|c}
\eb_j & (s_j, s_{j+1}, s_{j+2}) \\
\hline
1 0 0  & (1, 1, 1) \\
0 1 0  & (1, 0 ,1) \\
0 0 1  & (1, 0, 0)
\end{array}
$$
Since the three syndrome sequences are distinct,
we can map $(s_{j+1}, s_{j+2})$ to the corresponding single-error
pattern $\eb_j$ using a simple $3$-entry table lookup, and then correct
the error as indicated.  (If $(s_{j+1}, s_{j+2})$ is not in the table---
\ie if $(s_j, s_{j+1}, s_{j+2}) = (1, 1, 0)$--- then we have detected a
weight-2 error.)

We see that this simple algorithm can correct any single-error
pattern $\eb_j$, provided that there is no second error during blocks
$j$ through $j+2$.   The decoder synchronizes itself properly whenever a
zero syndrome occurs, and subsequently can correct one error in every
third block.

Finally, we consider tail-biting codes derived from the rate-$1/3$
convolutional code of Example 3.  In this case, it turns out that a
tail-biting interval of $N$ $3$-blocks results in no loss of minimum
distance whenever $N \ge 5$.

\pagebreak
\smallskip\noindent
\textbf{Example 4} (rate-$1/3$, single-error-correcting, CSS-type
tail-biting code).  Taking the rate-$2/3$ binary convolutional code
$\CC^\perp$ of Example 3 and a tail-biting interval of five $3$-blocks,
we obtain the following 10 tail-biting generators:
$$
\begin{array}{ccc|ccc|ccc|ccc|ccc}
1 & 1 & 0 & 0 & 1 & 1 & 0 & 0 & 0 & 0 & 0 & 0 & 0 & 0 & 0 \\
0 & 0 & 1 & 1 & 1 & 0 & 0 & 0 & 0 & 0 & 0 & 0 & 0 & 0 & 0 \\
0 & 0 & 0 & 1 & 1 & 0 & 0 & 1 & 1 & 0 & 0 & 0 & 0 & 0 & 0 \\
0 & 0 & 0 & 0 & 0 & 1 & 1 & 1 & 0 & 0 & 0 & 0 & 0 & 0 & 0 \\
0 & 0 & 0 & 0 & 0 & 0 & 1 & 1 & 0 & 0 & 1 & 1 & 0 & 0 & 0 \\
0 & 0 & 0 & 0 & 0 & 0 & 0 & 0 & 1 & 1 & 1 & 0 & 0 & 0 & 0 \\
0 & 0 & 0 & 0 & 0 & 0 & 0 & 0 & 0 & 1 & 1 & 0 & 0 & 1 & 1 \\
0 & 0 & 0 & 0 & 0 & 0 & 0 & 0 & 0 & 0 & 0 & 1 & 1 & 1 & 0 \\
0 & 1 & 1 & 0 & 0 & 0 & 0 & 0 & 0 & 0 & 0 & 0 & 1 & 1 & 0 \\
1 & 1 & 0 & 0 & 0 & 0 & 0 & 0 & 0 & 0 & 0 & 0 & 0 & 0 & 1 \\
\end{array}
$$
These generate a $(15,10,3)$ binary tail-biting code
$\B^\perp$ with minimum distance $d^\perp = 3$.
The orthogonal block code $\B$ is the corresponding $(15,5)$
binary tail-biting code $\B$ derived from $\CC$, which is necessarily
self-orthogonal.  Thus $\B$ specifies a $[15, 5, 3]$ CSS-type
single-error-correcting block stabilizer code.
\qed

To decode this code, we can use the same simple decoding algorithm as
for the corresponding convolutional code, but now on a circular time
axis.  If only a single error occurs, then the first syndrome 1 after
two zeroes (on a circular time axis) identifies the $3$-tuple block of 
the error, and the next two bits determine its position within the block,
according to the $3$-entry table above.  Indeed, the fact that these 15
syndrome $5$-tuples are distinct proves that single-error-correction is
possible, and thus that $\B^\perp$ has minimum distance 3.

\subsection{Comparison of single-error-correcting CSS-type codes}

We now compare our single-error-correcting CSS-type convolutional and
tail-biting codes with the single-error-correcting CSS-type
$7$-qubit Steane code (Example B).

For decoding error probability, we again
assume that the probability of an error in any qubit is
$p$, independent of errors in other qubits.  We do
not take into account that, because of the independence of the two
decoders, there are some weight-2 error patterns that are correctible
(\eg $X$ and $Z$);  this would yield a minor improvement (a factor of
7/9) in our estimates.

For the $7$-qubit Steane code, a decoding error may occur if
there are 2 errors in any block, so the error probability is of
the order of ${7 \choose 2} p^2 = 21p^2$ per block, or per encoded
qubit.

For our rate-$1/3$ convolutional code, for each $3$-block, a
decoding error may occur if there are two errors in that $3$-block, or 
one in that $3$-block and one in the two subsequent $3$-blocks.  The error
probability is therefore of the order of $(3 + 3\cdot 6) p^2 = 21 p^2$
per $3$-block, or per encoded qubit.

Lastly, for our $[15, 5, 3]$ tail-biting code, a decoding error
requires 2 errors in a block of 15 qubits, so the error
probability is of the  order of ${15 \choose 2} p^2 = 105p^2$ per block,
or $21p^2$ per encoded qubit.

Again, we conclude that the decoding error probability is very nearly
the same for any of these codes, and is about twice that of the
$\F_4$-linear codes considered earlier.

With regard to rate and decoding complexity, our CSS-type convolutional
and tail-biting codes again have rate $1/3$, greater than that of
any previous simple CSS-type single-error-correcting code.  The decoder
for Examples 3 and 4 only uses a $3$-entry table lookup twice, and is
arguably even simpler than the simple decoder for Examples 1 and 2.

Our convolutional code rate and error-correction capability (one error
every three $3$-blocks) are comparable to those of a $[9, 3, 3]$ 
CSS-type block code.  However,  it can be shown (by linear programming) that
no $[9, 3, 3]$ CSS-type block code exists \cite{Gra02,Gra}.

Our tail-biting code is a $[15,5,3]$ CSS-type block code.  We could
obtain a code with the same parameters by shortening a
$[31, 21, 3]$ CSS-type block code.  However, such a shortened code would
not necessarily have such a simple structure as our tail-biting
code, nor such a simple decoding algorithm.

\section{Algebraic theory of $\F_4$-linear convolutional codes}

In this section we give a brief presentation of the algebraic theory
of $\F_4$-linear convolutional codes.  We focus on those results which
are most helpful in searching for good codes and for analyzing their
performance.  The theory of $\F_2$-linear convolutional codes is
analogous.  For further background, see \cite{F70}, or any text that
covers convolutional codes, such as \cite{JZ99}.

\subsection{Rate-$1/n$ convolutional codes}

We have defined a rate-$1/n$ $\F_4$-linear shift-invariant
convolutional code $\CC$ with constraint length $\nu$ as the set of all
$\F_4$-linear combinations of the shifts of a single basic finite
generator sequence $\gb = \{g_{jk}, 1 \le j \le n,$ $0 \le k \le \nu\}$
by an integral number of $n$-blocks.

It is helpful to use ``$D$-transform" (generating function)
notation, as is standard for convolutional codes.  A sequence of
$n$-blocks such as $\gb$ is written as an $n$-tuple $\gb(D) = \{g_j(D),
1 \le j \le n\}$ of $D$-transforms $g_j(D) = \sum_k g_{jk}D^k$, where $D$
is an indeterminate, called the \emph{shift operator}.
For example, the generator  $\gb
= (\ldots, 000, 111, 1\omega\overline{\omega}, 000, \ldots)$ is written
as the polynomial $3$-tuple
$$
\gb(D) = (g_1(D), g_2(D), g_3(D)) = (1 + D, 1 + \omega D, 1 +
\overline{\omega}D),
$$
where we have aligned the first nonzero block with the index $k = 0$.

A shift of $\gb(D)$ by an integral number $\ell$ of
$n$-blocks is then represented by the $D$-transform $D^\ell \gb(D) =
\{D^\ell g_j(D), 1 \le j \le n\}$, where
$D^\ell g_j(D) = \sum_{k} g_{jk}D^{k+\ell}$.  For example, a one-block
shift of the generator $\gb(D)$ is represented by
$D\gb(D) = (D + D^2, D + \omega D^2, D + \overline{\omega}D^2)$.

The rate-$1/n$ convolutional code $\CC$ is then the set of ``all"
$\F_4$-linear combinations of the shifted generators $\{D^\ell \gb(D),
\ell \in \Z\}$;  \ie
$$
\CC = \{\sum_{\ell\in \Z} u_\ell D^\ell \gb(D) \mid u_\ell \in \F_4\} =
\{u(D) \gb(D)\},
$$
where we have defined $u(D)$ as the $D$-transform $\sum_\ell u_\ell
D^\ell$ of the coefficient sequence $\{u_\ell\}$.

We have put ``all" in quotation marks
because usually, for technical reasons, the coefficient sequence
$\{u_\ell\}$ is required to have only finitely many nonzero $u_\ell$
with negative indices
$\ell < 0$.  Such a sequence is called a \emph{Laurent power series},
and
the set of all Laurent power series in $D$ over $\F_4$ is denoted by
$\F_4((D))$.  In short,
$$
\CC = \{u(D)\gb(D) \mid u(D) \in \F_4((D))\}.
$$

The set $\F_4((D))$ is shift-invariant; \ie $D^\ell\F_4((D)) =
\F_4((D))$ for any $\ell \in \Z$.  Consequently, $\CC$ is shift-invariant:
$D^\ell\CC = \CC$.

The set $\F_4((D))$ is actually a field under sequence
(componentwise) addition and sequence (polynomial) multiplication
(\ie $a(D)b(D) = \sum_k D^k \sum_{k'} a_{k'} b_{k-k'}$).  In particular,
every nonzero sequence
$u(D) \in \F_4((D))$ has a multiplicative inverse $1/u(D)$, which may be
found by polynomial long division.  For example, the inverse of $1+D$ is
$1 + D + D^2 + \cdots$.

The set $\F_4((D))^n$ of all $n$-tuples of Laurent power series is an
$n$-dimensional vector space over the field $\F_4((D))$.  A rate-$1/n$
code $\CC$ is therefore simply a one-dimensional shift-invariant
subspace of $\F_4((D))^n$.  Any nonzero code sequence $u(D)\gb(D) \in \CC$
may thus be taken as a generator for $\CC$.

We wish to choose a canonical generator $\gb(D) \in \CC$ that has
the most desirable properties.  For most purposes, the best choice is
a polynomial code sequence $\gb(D) \in \CC$ of least degree, where we
define $\deg \gb(D) = \max_j \{\deg g_j(D)\}$.  Given any nonzero
polynomial code sequence $\cb(D) = (c_1(D), \ldots, c_n(D)) \in \CC$,
a minimum-degree polynomial generator is $\gb(D) = \cb(D)/d(D)$, where
$d(D)$ is the greatest common divisor of the polynomials $c_j(D)$.
    Conversely, a polynomial generator $\gb(D)$ is
minimum-degree if and only if its components $g_j(D)$ are relatively
prime.  The minimum-degree polynomial generator $\gb(D)$ is thus
unique up to multiplication by nonzero scalars in $\F_4$.

     A minimum-degree polynomial generator $\gb(D)$ has the following
properties \cite{F70, JZ99}:
\begin{itemize}
\item A code sequence $\cb(D) = u(D)\gb(D) \in \CC$ is polynomial if and
only if
$u(D)$ is polynomial; \ie the set
of all polynomial code sequences is $\{u(D)\gb(D), u(D)
\in \F_4[D]\}$, where $\F_4[D]$ denotes the set of all polynomials in
$D$ over $\F_4$.  This is called the
\emph{noncatastrophic property}.
\item The constraint length $\nu = \deg \gb(D)$ is minimized.
\end{itemize}

\subsection{Orthogonality}

The \emph{Hermitian inner product} of two Laurent power series $a(D),
b(D) \in \F_4((D))$ is defined as
$$
\inner{a(D)}{b(D)} = \sum_{k \in \Z} a_{k}^\dagger b_{k}.
$$
The sum is well defined if and only if there are only finitely many
nonzero summands  $a_{k}^\dagger b_{k}$.

The Hermitian inner product of two Laurent $n$-tuples $\ab(D),
\bb(D) \in \F_4((D))^n$ is defined as
$$
\inner{\ab(D)}{\bb(D)} = \sum_{j=1}^n \inner{a_j(D)}{b_j(D)}.
$$

The \emph{cross-correlation sequence} of
$\ab(D), \bb(D)$ is defined as
$$
R_{\ab \bb}(D) = \sum_{j=1}^n a_j^\dagger(D^{-1}) b_j(D),
$$
again assuming well-defined products $a_j^\dagger(D^{-1}) b_j(D)$.
Thus 

$$R_{\ab \bb,\ell} = \sum_j \sum_{k} a_{jk}^\dagger b_{j,k-\ell}
= \inner{\ab(D)}{D^\ell\bb(D)}.$$  
Therefore a sequence $\ab(D)$
is orthogonal to all shifts $D^\ell\bb(D)$ of a sequence $\bb(D)$ if
and only if
$$
R_{\ab \bb}(D) = 0;
$$
\ie if and only if all cross-correlation terms $R_{\ab \bb,\ell}$ are
equal to zero. \pagebreak

A rate-$1/n$ linear shift-invariant convolutional code
$\CC$ with generator $\gb(D)$ is thus self-orthogonal if and only if
$$R_{\gb\gb}(D) = \sum_{j=1}^n g_j^\dagger(D^{-1}) g_j(D) = 0.$$

\smallskip\noindent
\textbf{Example 1}.
The sequence $\gb(D) = (1 + D, 1 + \omega D, 1 +
\overline{\omega}D)$ is orthogonal to all of its shifts since
\begin{eqnarray*}
R_{\gb \gb}(D) & = & (1 + D^{-1})(1 + D) +
(1 + \overline{\omega}D^{-1})(1 + \omega D) +
(1 + \omega D^{-1})(1 + \overline{\omega} D) \\
& = & (D^{-1} + D) + (\overline{\omega}D^{-1} + \omega D) + (\omega
D^{-1} + \overline{\omega} D) = 0.
\end{eqnarray*}
Thus the convolutional code $\CC$ generated by all shifts of
$\gb(D)$ is self-orthogonal.  \qed

\smallskip\noindent
\textbf{Example 3}.
The sequence $\gb'(D) = (1 + D + D^2, 1 + D^2, 1)$ is orthogonal to all
of its shifts since
\begin{eqnarray*}
R_{\gb' \gb'}(D) & = & (1 + D^{-1} + D^{-2})(1 + D + D^2) +
(1 + D^{-2})(1 + D^2) + 1 \\
& = & (D^{-2} + 1 + D^2) + (D^{-2} + D^2) + 1 = 0.
\end{eqnarray*}
Thus the convolutional code $\CC'$ generated by all shifts of
$\gb'(D)$ is self-orthogonal.  \qed

The orthogonal code $\CC^\perp$ to a rate-$1/n$ convolutional code $\CC$
with generator $\gb(D)$ is the set of all sequences $\ab(D)$
that are orthogonal to all shifts $D^k\gb(D)$, and thus are
uncorrelated with $\gb(D)$--- \ie such that $R_{\ab \gb}(D) = 0$.
It follows that $\CC^\perp$ is a rate-$(n-1)/n$
$\F_4$-linear shift-invariant convolutional code--- \ie an
$(n-1)$-dimensional shift-invariant subspace of $\F_4((D))^n$.

It is again desirable to choose as generators for $\CC^\perp$ a set of
$n-1$ linearly independent minimum-degree polynomial generators
$\hb_i(D), 1 \le i \le n-1$, such that $R_{\hb_i \gb}(D) = 0$.  This
can be done by an exhaustive search for low-degree orthogonal
polynomial sequences, or by various algebraic methods.  Again, such a
minimal-degree generator set has the following properties \cite{F70,
JZ99}:
\begin{itemize}
\item A code sequence $\cb(D) = \sum_i u_i(D)\hb_i(D) \in \CC^\perp$ is
polynomial if and only if $\ub(D)$ is polynomial; \ie the generator set
is \emph{noncatastrophic}.
\item The total constraint length $\nu^\perp = \sum_i \nu_i^\perp =
\sum_i \deg \hb_i(D)$ of the generator set is minimized, and is equal to
the constraint length
$\nu$ of $\CC$ \cite{F73}.   This latter property may be used to check
whether a set of independent orthogonal generators is a minimal-degree set.
\end{itemize}

\smallskip\noindent
\textbf{Example 1}.
For the code $\CC$ generated by the degree-1 generator $\gb(D) = (1 +
D, 1 + \omega D, 1 + \overline{\omega}D)$, the two sequences $\hb_1(D) =
\gb(D)$ and $\hb_2(D) = (\overline{\omega},
\omega, 1)$   are independent, are orthogonal to
$\gb(D)$, and have degrees $\nu_1^\perp = 1$ and
$\nu_2^\perp = 0$ that sum to $\nu = 1$.  Therefore
$\{\hb_1(D), \hb_2(D)\}$ is a minimal-degree set of generators for the
orthogonal rate-$2/3$ code $\CC^\perp$.  \qed

\smallskip\noindent
\textbf{Example 3}.
For the code $\CC'$ generated by the degree-2 generator $\gb'(D) = (1 +
D + D^2, 1 + D^2, 1)$, the two sequences $\hb'_1(D) =
(1, 1+D, D)$ and $\hb'_2(D) = (D, D, 1)$ are independent,
are orthogonal to $\gb'(D)$, and have degrees $\nu_1^\perp = 1$ and
$\nu_2^\perp = 1$ that sum to $\nu = 2$.  Therefore
$\{\hb'_1(D), \hb'_2(D)\}$ is a minimal-degree set of generators for the
orthogonal rate-$2/3$ code ${\CC'}^\perp$.  \qed

In view of the noncatastrophic property, the
minimum Hamming distance $d^\perp$ of the orthogonal code $\CC^\perp$ is
the minimum weight of any nonzero polynomial code sequence $\sum_i
u_i(D)
\hb_i(D)$, where $\{u_i(D), 1 \le i \le n-1\}$ is any set of
polynomial sequences.  For simple codes, the minimum-weight nonzero
sequence will be a polynomial sequence of low degree, and will often be
obvious by inspection.  For instance, for both of our example codes, one
generator $\hb_i(D)$ has weight 3, and it is easy to see that no
nonzero sequence in $\CC^\perp$ can have weight less than 3, so
$d^\perp = 3$.

\subsection{Convolutional code symmetries}

In searching for generators of good codes, it is helpful to observe that
there are certain symmetries that preserve the most important properties
of convolutional codes.  A symmetry that converts $\gb(D), \hb(D) \in
\F_4((D))^n$ to $\gb'(D), \hb'(D) \in \F_4((D))^n$ will be called
\emph{weight-preserving} if Hamming weights are preserved, and
\emph{orthogonality-preserving} if $R_{\gb\hb}(D) = 0$ implies
$R_{\gb'\hb'}(D) = 0$.

\begin{theorem}[Convolutional code symmetries]
The following symmetries of $\F_4((D))^n$ are both weight-preserving and
orthogonality-preserving:
\begin{enumerate}
\item Multiplication of any component $g_j(D)$ by any monomial $\alpha
D^\ell$, $\alpha \neq 0, \ell \in \Z$.
\item Conjugation:  $\gb(D) \to \gb^\dagger(D)$.
\item Time-reversal:  $\gb(D) \to \gb(D^{-1})$.
\item Modulation: $\gb(D) \to \gb(\alpha D)$ for any nonzero scalar
$\alpha \in \F_4$.
\item Permutation of the components $g_j(D)$.
\end{enumerate}
\end{theorem}

\emph{Proof}.  It is obvious that each symmetry is
weight-preserving.

To show that each symmetry is
orthogonality-preserving, we argue in each case that if
$R_{\gb\hb}(D) = 0$ and $\gb(D)$ and $\hb(D)$ are changed to
$\gb'(D)$ and $\hb'(D)$, then
$R_{\gb'\hb'}(D) = 0$, using
$$R_{\gb \hb}(D) = \sum_{j=1}^n g_j^\dagger(D^{-1}) h_j(D).$$
\begin{enumerate}
\item If $g_j(D), h_j(D) \to \alpha D^\ell g_j(D), \alpha D^\ell
h_j(D)$, then $R_{\gb'\hb'}(D) = R_{\gb\hb}(D)$, since \\
$\alpha^\dagger D^{-\ell} g_j^\dagger(D^{-1}) \alpha D^{\ell} h_j(D) =
g_j^\dagger(D^{-1})h_j(D)$.
\item If  $\gb(D), \hb(D) \to \gb^\dagger(D), \hb^\dagger(D)$, then
$R_{\gb\hb}(D) \to R^\dagger_{\gb\hb}(D)$.
\item If $\gb(D), \hb(D) \to \gb(D^{-1}), \hb(D^{-1})$, then
$R_{\gb\hb}(D) \to R_{\gb\hb}(D^{-1})$.
\item If $\gb(D), \hb(D) \to \gb(\alpha D), \hb(\alpha D)$, then
$R_{\gb\hb}(D) \to R_{\gb\hb}(\alpha D)$.
\item Permutation of the $g_j(D)$ and $h_j(D)$ in the same way does not
affect $R_{\gb\hb}(D)$.  \qed
\end{enumerate}

In particular, if $\gb(D)$ is a self-orthogonal generator, then the
modified generator $\gb'(D)$ under any of these symmetries is
self-orthogonal.

The first symmetry shows that, without loss of generality, we may assume
that all component generators $g_j(D)$ are \emph{monic} polynomials;
\ie the zero-degree coefficient $g_{j,0}$ of $g_j(D)$ is  1.

\smallskip\noindent
\textbf{Example 1}.  It is easy to see that a degree-1 generator is
self-orthogonal if and only if it is equivalent to  $\gb(D) = (1
+ D, 1 + \omega D, 1 +
\overline{\omega}D)$ under one of these symmetries (see Section V).
\qed

\smallskip\noindent
\textbf{Example 3}.
The degree-2 binary generator $\gb'(D) = (1 + D + D^2, 1 + D^2, 1)$ is
invariant under conjugation, and effectively invariant under
time-reversal.  There are 6 equivalent binary generators under
component permutations.  No further equivalent binary generators
are produced by the symmetry $\gb(D) \to \gb(\alpha D)$.
As we will see in Section V,
$\gb'(D)$ is the unique monic degree-2 binary self-orthogonal generator,
up to component permutations.
\qed

\smallskip\noindent
\textbf{Example 5}.  The degree-2 generator $\gb''(D) = (1+D+D^2, 1
+ \omega D + D^2, 1+D)$ satisfies $R_{\gb''\gb''}(D) = 0$, so the
convolutional code $\CC''$ generated by all shifts of
$\gb''(D)$ is self-orthogonal (see Section VI).  A minimal-degree
generator set for the orthogonal code $(\CC'')^\perp$ is $\{\hb_1(D) =
(\omega D,
\overline{\omega} D, 1+D), \hb_2(D) = (1, 1 + \overline{\omega} D, 1 +
\overline{\omega} D)\}$.  The minimum distance of  $(\CC'')^\perp$ is
$d^\perp = 4$. There are 6 equivalent generators to   $\gb''(D)
= (1+D+D^2, 1  + \omega D + D^2, 1+D)$ under conjugation and the
symmetry $\gb(D) \to \gb(\alpha D)$, or 36 if component permutations are
also considered.
\qed

\section{Rate-$1/n$ single-error-correcting codes}

Using the theoretical development of Section IV, it is straightforward
to find all possible short-constraint-length, single-error-correcting,
convolutional stabilizer codes based on both binary and $\F_4$-linear
rate-$1/n$ convolutional codes.

In order that a rate-$1/n$ linear shift-invariant convolutional code
$\CC$ generated by $\gb(D) = (g_1(D), \ldots, g_n(D))$ has an orthogonal
code $\CC^\perp$ with minimum distance $d^\perp \ge 3$, it is necessary
and sufficient that all component generator polynomials $g_j(D)$ be
linearly independent, so that no weight-2 error pattern can cause a zero
syndrome.  If all generator polynomials $g_j(D)$ are restricted to be
monic, then this reduces to the requirement that all $g_j(D)$ be
different.

To find single-error-correcting stabilizer codes, it therefore suffices
to list all monic polynomials $g(D)$ of low degree, with their
autocorrelation functions $R_{gg}(D) = g^\dagger(D^{-1}) g(D)$, and to
identify all subsets $\gb(D)$ of size $n$ such that $R_{\gb\gb}(D) =
\sum_{j=1}^n R_{g_j g_j}(D) = 0$.

In Table I, we therefore list all binary polynomials $g(D)$ of degree 3
or less, with the non-negative-degree components $[R_{gg}(D)]_{0^+}$ of
their autocorrelation functions (the negative-degree components are
symmetric).  For $3 \le n \le 8$, subsets of size $n$ are identified
such that the corresponding autocorrelation functions sum to
zero.  There exists a
unique binary self-orthogonal rate-$1/n$ convolutional code with
constraint length $\nu = 2$:  namely, the rate-$1/3$ Example 3 code.
Seven further codes are listed with constraint length $\nu = 3$ and
rates
from 1/4 down to 1/8. (It is easy to
verify that none of these generator sets is catastrophic.)  In turn,
these binary codes yield single-error-correcting CSS-type convolutional
stabilizer codes with quantum code rates ranging from $1/3$ up to $6/8$.

\begin{figure}[hbt]
$$
\begin{array}{|l|l|cccccccc|}
\hline
\rule[-7pt]{0pt}{20pt} %MG add some space for the fractions
g(D) & [R_{gg}(D)]_{0^+} & \frac{1}{3} & \frac{1}{4} & \frac{1}{4} &
\frac{1}{5} & \frac{1}{5} &  \frac{1}{5} & \frac{1}{6} & \frac{1}{8} \\
\hline
1 & 1 & * & * & & * & * &  & * & * \\
1 + D & D & & & * & & * & * & * & * \\
1 + D^2 & D^2 & * & * & & * & * &  & * & * \\
1 + D + D^2 & 1 + D^2 & * & & * & * & & & * & * \\
1 + D^3 & D^3 & & & * & & * & * & * & * \\
1 + D + D^3 & 1 + D + D^2 + D^3 & & * & & * & * & * & & * \\
1 + D^2 + D^3 & 1 + D + D^2 + D^3 & & & * & * & & * & & * \\
1 + D + D^2 + D^3 & D + D^3 & & * & & & & * & * & * \\
\hline
\end{array}
$$
\begin{center}
Table I.  Self-orthogonal binary rate-$1/n$ convolutional codes.
\end{center}
\end{figure}

To decode these rate-$1/n$ CSS-type codes, $n \ge 4$, as with our
rate-$1/3$ CSS-type code, bit flip and phase flip bits may be decoded
independently in two binary decoders.  Since there are only $n$ possible
single-error patterns  $\eb_j$ in the $j$th $n$-block, decoding requires
only an $n$-entry table lookup.  Decoding will succeed if there is no
second error during blocks $j$ through $j+3$;  \ie each decoder can
correct 1 error in every 4 $n$-blocks.

     The minimum-length single-error-correcting tail-biting code that
can be derived from any of these codes is easily determined by finding
the minimum tail-biting length for which all cyclic shifts of all $n$
single-error syndromes are distinct.  For the eight codes listed in
Table I, the minimum-length corresponding tail-biting codes are listed
in Table II. Additionally, we give the number $N_{d^\perp}$ of 
words of weight $d^\perp$ in $\B^\perp$, and an upper bound
$d_{\mathrm{CSS}}$ on the minimum distance of a CSS-type code.
   Again, these codes may be decoded by the same simple
$n$-entry table lookup algorithm, operating on a circular time
axis.

\begin{figure}[hbt]
$$
\begin{array}{|c|l|c|c|c|c|c|c|c|}
\hline
\rule[-7pt]{0pt}{20pt} %MG add some space for the fractions
\mathrm{rate} & \nu & N_{d^\perp} &\B & \B^\perp & \mathrm{stabilizer
~code} & d_{\mathrm{CSS}} %& d_{\mathrm{opt}}
\\[1ex] % \alpha&
\hline
1/3 &2 &  2& (15, 5, 6)  & (15, 10, 3)  & [15,  5, 3] & 3 %& 4
\\ %  1/2 &
1/4 &3 &  4& (20, 5, 8)  & (20, 15, 3)  & [20, 10, 3] & 3 %& 4
\\ %  1/2 &
1/4 &3 &  2& (20, 5, 8)  & (20, 15, 3)  & [20, 10, 3] & 3 %& 4
\\ %  1/3 &
1/5 &3 &  6& (30, 6, 12) & (30, 24, 3)  & [30, 18, 3] & 3 %& 4
\\ %  2/5 &
1/5 &3 &  9& (35, 7, 10) & (35, 28, 3)  & [35, 21, 3] & \Red{3\mbox{--}4} %&4\mbox{--}5
\\ %  1/3 &
1/5 &3 &  3& (35, 7, 14) & (35, 28, 3)  & [35,21, 3] & \Red{3\mbox{--}4} %&4\mbox{--}5
\\ %  1/3 &
1/6 &3 & 15& (42, 7, 14) & (42, 35,3)  & [42, 28, 3] & 3\mbox{--}4
%& 4\mbox{--}5
\\ %  1/3 &
1/8 &3 & 28& (56, 7, 20) & (56, 49, 3)  & [56, 42, 3] & 3\mbox{--}4
%& 4\mbox{--}5
\\
%  1/3 &
\hline
\end{array}
$$
\begin{center}
Table II.  CSS-type rate-$1/n$ tail-biting codes.
\end{center}
\end{figure}

Similarly, in Table III we list all 16 monic quaternary polynomials
$g(D)$ of degree 2 or less, along with the non-negative-degree
components
$[R_{gg}(D)]_{0^+}$ of their autocorrelation functions.  For $3 \le n
\le 16$, certain subsets of size $n$ are identified such that the sums
of
the corresponding autocorrelation functions are zero.  There exists a
unique $\F_4$-linear self-orthogonal rate-$1/n$ convolutional code with
constraint length $\nu = 1$:  namely, the rate-$1/3$ Example 1 code.
Five further codes are listed with constraint length $\nu = 2$ and rates
$1/4$, $1/5$, $1/6$, $1/10$ and $1/16$. (Again, none of these generator
sets is catastrophic.)  These codes yield single-error-correcting
convolutional stabilizer codes with quantum code rates ranging from
$1/3$ up to
$14/16$.

\begin{figure}[hbt]
$$
\begin{array}{|l|l|cccccc|}
\hline
\rule[-7pt]{0pt}{20pt} %MG add some space for the fractions
g(D) & [R_{gg}(D)]_{0^+} & \frac{1}{3} & \frac{1}{4} &
\frac{1}{5} & \frac{1}{6} &  \frac{1}{10} & \frac{1}{16} \\
\hline
1 & 1 & & * & & & * & * \\
1 + D & D & * & * & * & * & & * \\
1 + \omega D & \omega D & * & & * & * & & * \\
1 + \overline{\omega} D & \overline{\omega} D & * & & * & * & & * \\
1 + D^2 & D^2 & & * & & * & & * \\
1 + \omega D^2 & \omega D^2 & & & & * & & * \\
1 + \overline{\omega} D^2 & \overline{\omega} D^2 & & & & * & & * \\
1 + D + D^2 & 1 + D^2 & & & & & * & * \\
1 + D + \omega D^2 & 1 + \overline{\omega} D + \omega D^2 & & & & & * &
*
\\
1 + D + \overline{\omega} D^2 & 1 + \omega D + \overline{\omega} D^2 & &
& & & * & * \\
1 + \omega D + D^2 & 1 + D + D^2 & & * & * & & * & * \\
1 + \omega D + \omega D^2 & 1 + \overline{\omega} D + \omega D^2 & & & &
& * & * \\
1 + \omega D + \overline{\omega} D^2 & 1 + \overline{\omega} D^2 & & & &
& * & * \\
1 + \overline{\omega} D + D^2 & 1 + D + D^2 & & & * & & * & * \\
1 + \overline{\omega} D + \omega D^2 & 1 + \omega
D^2 & & & & & * & * \\
1 + \overline{\omega} D + \overline{\omega} D^2 & 1 + \omega D +
\overline{\omega} D^2 & & & & & * & * \\
\hline
\end{array}
$$
\begin{center}
Table III.  Self-orthogonal $\F_4$-linear rate-$1/n$ convolutional
codes.
\end{center}
\end{figure}

In this case, since there are only $3n$ possible quaternary
single-error patterns  $\Eb_j$ in the $j$th $n$-block, decoding requires
only a $3n$-entry table lookup.  Decoding will succeed if there is no
second error during blocks $j$ through $j+2$;  \ie each decoder can
correct 1 error in every 3 $n$-blocks.

Again, the minimum-length corresponding single-error-correcting
tail-biting code may be
determined by finding the minimum tail-biting length for which all
cyclic shifts of all $3n$ single-error syndromes are distinct.  For the
six codes listed in Table III, these tail-biting codes are listed in
Table IV.
Additionally, we give the number $N_{d^\perp}$ of words of
weight $d^\perp$ in $\B^\perp$, and an upper bound
  $d_{\mathrm{opt}}$
on the minimum distance of a general quantum code.
  Most of the codes meet this bound on
minimum distance.
  Again, these codes may be decoded by the same simple
$3n$-entry table lookup algorithm, operating on a circular time
axis.

%\begin{figure}[hbt]
$$
\begin{array}{|l|c|c|c|c|c|c|}
\hline
\rule[-7pt]{0pt}{20pt} %MG add some space for the fractions
\mathrm{rate} & \nu & N_{d^\perp}&\B & \B^\perp & \mathrm{stabilizer
~code} & d_{\mathrm{opt}} \\ %& \alpha
\hline
1/3  & 1 & 3  &(9, 3)  & (9, 6, 3)   & [9, 3, 3]   & 3\\  % & 1/2
1/4  & 2 & 12 &(20, 5) & (20, 15, 3) & [20, 10, 3] & 4\\  % & 1/4
1/5  & 2 & 15 &(15, 3) & (15, 12, 3) & [15, 9, 3]  & 3\\  % & 4/12
1/6  & 2 & 33 &(30, 5) & (30, 25, 3) & [30, 20, 3] & 4\\  % & 1/4
1/10 & 2 &108 &(40, 4) & (40, 36, 3) & [40, 32, 3] & 3\\
1/16 & 2 &600 &(80, 5) & (80, 75, 3) & [80, 70, 3] & 3\mbox{--}4\\
\hline
\end{array}
$$
\begin{center}
Table IV.  $\F_4$-linear rate-$1/n$ tail-biting codes.
\end{center}
%\end{figure}

\section{Rate-$1/3$ codes with $d > 3$}

We have performed a computer search for both binary and $\F_4$-linear
self-orthogonal convolutional codes with constraint lengths up to $\nu
= \Red{12}$ and $\nu = \Red{6}$, respectively.  The best codes found
have orthogonal codes with Hamming distances $d^\perp = \Red{10}$ and
$d^\perp = \Red{9}$, respectively. 

We examined one code from each equivalence class
under the symmetries of Theorem 1.  In particular, we considered only
monic generators $\gb(D)$.  We eliminated catastrophic generators.  For
each code found, we found a minimal-degree pair of orthogonal
generators, $\hb_1(D)$ and
$\hb_2(D)$, such that $\deg \hb_1(D) + \deg \hb_2(D) = \deg \gb(D)$.  We
then found the minimum distance of the orthogonal code by a trellis
search.  For the
$\F_4$-linear codes, we found the notation of
J\"{o}nsson \cite{J03} to be helpful.

Table V shows the best binary codes found for constraint lengths $2
\le \nu \le \Red{12}$.  The best code is the one whose orthogonal code
has the greatest minimum distance $d^\perp$; to resolve ties,
minimization of the number $N_{d^\perp}$ of code sequences of weight
$d^\perp$
%(and hence the union bound on the error probability)
is used as
a secondary criterion.  \Red{A unique best code (up to the symmetries
of Theorem 1) was found for $\nu = 2, 3, 5, 6,7,10,11$ and $12$.}
For brevity, we represent polynomials by their coefficient sequences;
\eg $1101 = 1 + D + D^3$.

%\begin{figure}[hbt]
$$\def\arraystretch{0.9}
\begin{array}{|c|l@{\,}l@{\,}l|l@{\,}l@{\,}l|c|c|}
\hline
\rule[-7pt]{0pt}{20pt} %MG add some space for the fractions
\nu &
\multicolumn{3}{c|}{\gb(D)} &
\multicolumn{3}{c|}{\hb_1(D), \hb_2(D)}&
%\multicolumn{3}{c|}{\hb_2(D)}&
d^\perp & N_{d^\perp}\\
\hline
  2&      1&              101&            111&         11&10&11          
       & 3& 2\\
   &       &                 &               &         01&11&10          
       &  &  \\
\hline
  3&      111&            1101&           1111&        101&101&100       
       & 4& 3\\
   &         &                &               &        01&10&11          
       &  &  \\
\hline
  4&      1111&           11001&          10101&       101&010&001       
       & 4& 1\\
   &          &                &               &       111&111&100       
       &  &  \\
  4&      1101&           10011&          11011&       0001&1000&1001    
       & 4& 1\\
   &          &                &               &       11&11&10          
       &  &  \\
  4&      1101&           11001&          11011&       1001&1001&1000    
       & 4& 1\\
   &       &                 &               &         01&10&11          
       &  &  \\
\hline
  5&      11111&          101101&         101111&      1011&0001&0100    
       & 5& 1\\
   &           &                &               &      111&110&111       
       &  &  \\
\hline
  6&      111001&         1100111&        1001111&     1111&1000&1101    
       & 6& 2\\
   &            &                &               &     1101&0011&0110    
       &  &  \\
\hline
  7&      1010001&        11110101&       11100011&    10111&00001&01100 
       & 7& 7\\
   &             &                &               &    1101&1110&1101    
       &  &  \\
\hline
  8&      11010101&       110100101&      111111011&   10101&11010&11011 
       & 7& 1\\
   &              &                &               &   11101&01001&00100 
       &  &  \\
  8&      11001001&       111000101&      100110101&   01011&11000&11101 
       & 7& 1\\
   &              &                &               &   11011&00111&01000 
       &  &  \\
  8&      10100001&       111011101&      110111111&   00111&11010&10011 
       & 7& 1\\
   &              &                &               &   10001&11011&11100 
       &  &  \\
  8&      10110001&       111110011&      101101111&   
100101&000011&011000    & 7& 1\\
   &              &                &               &   1001&1110&1011    
       &  &  \\
\hline
  9&      101000001&      1100111101&     1110011111&  
110111&101001&100000    & 8& 3\\
   &               &                &               &  11101&00110&01101 
       &  &  \\
  9&      111011011&      1011000001&     1000111111&  
111111&101100&100001    & 8& 3\\
   &               &                &               &  10011&01111&00010 
       &  &  \\
\hline
10&      10111110101&    11110101001&    10101110110& 
1011111&1011000&0101111 & 9& 8\\
   &                 &               &               & 11001&10111&11000 
       &  &  \\
\hline
11&      100001010111&   110010101011&   
101110000010&1110011&1011101&1010100 & 9& 1\\
   &                  &               &               
&010011&000110&110101    &  &  \\
\hline\Red{
12&     1110010000010&  1101110010011&  1011111000111&0111111&1001010&1000111 & 10 & 5\\
  &                  &               &                &1010001&1111001&1010100&    &  }\\
\hline
\end{array}
$$
\begin{center}
Table V.  Self-orthogonal binary linear rate-$1/3$ convolutional codes.
\end{center}
%\end{figure}

Similarly, Table VI shows the best $\F_4$-linear codes found for
constraint lengths $1 \le \nu \le 6$.  \Red{For $1\le\nu\le 5$, the codes 
are unique up to equivalence.}

%\begin{figure}[hbt]
$$\def\w{\omega}\def\ww{\overline{\omega}}
\begin{array}{|c|l@{\,}l@{\,}l|@{\,}l@{\,}l@{\,}l|l@{\,}l@{\,}l|c|c|c}
\hline
\rule[-7pt]{0pt}{20pt} %MG add some space for the fractions
\nu &
\multicolumn{3}{c|}{\gb(D)} &
\multicolumn{3}{c|}{\hb_1(D)}&
\multicolumn{3}{c|}{\hb_2(D)}&
d^\perp & N_{d^\perp}\\
\hline
1 & 11& 1\omega& 1\omegabar & \omegabar& \omega& 1  & 11& 1\omega&
1\omegabar & 3 & 3 \\
2 & 111& 1\omega 1& 110 & 0\omega& 0\omegabar& 11 & 10& 1\omegabar&
1\omegabar & 4 & 12 \\
3 & 1001& 111\omegabar& 1\omega\omegabar\omega & 10& 1\omegabar&
1\omegabar & \omega 0\omega& 1\omegabar 1& 00\omegabar & 5 & 3 \\
% MG:
%  wrong generators for the dual code
%
%  4 & 1\omega\omegabar\omegabar 1& 1\omegabar 01\omegabar& 
% 111\omega\omega
%  & 1\omegabar\omega& \omegabar 01& \omega 0\omegabar & \omega 11&
%  \omegabar\omegabar 1& 0\omegabar 0 & 6 & 3 \\
4 & 1\omega\omegabar\omegabar 1& 1\omegabar 01\omegabar& 
111\omega\omega &
\omegabar\omega1&\omegabar\omega&\omegabar\omega1 &
\omegabar\omegabar1&\omega11&0\omegabar & 6 & 3 \\
5 & 11\omega 0\omegabar 1& 11\omegabar 10\omegabar&
1\omegabar\omega\omega\omega\omega & \omegabar\omega 1& 10\omegabar&
\omega\omegabar\omega & \omegabar 10\omegabar& \omega 0\omegabar 1&
00\omega\omega & 8 & 75 \\
\hline\Red{
6 & 1\ww\w1\w0\w&11\ww00\ww\ww&100\w1\ww1   & \ww\w11&11\ww&0101    & \w\w\w1&10\w1&\ww\w\ww & 9 & 78\\
6 & 1\ww\w1\w0\ww&1\w0\w\ww\w\w&11\w0\w\ww1 & \w\ww\w1&0\w1&\ww101  & 1\w01&\ww1\w1&0\w\ww   & 9 & 78\\
6 & 1\w1\ww\ww0\ww&1\ww\w\ww\ww11&1001\w1\w & \w\ww11&\ww\ww&0\w\w1 & 10\ww1&0111&\ww\ww\ww  & 9 & 78\\
6 & 11\w110\ww&10\w\w0\w\w&1\ww1\ww\w\ww1   & 1\w\w1&\ww0\ww&01\w1  & \w111&01\ww1&\ww0\ww   & 9 & 78}\\
\hline
\end{array}
$$
\begin{center}
Table VI.  Self-orthogonal $\F_4$-linear rate-$1/3$ convolutional codes.
\end{center}
%\end{figure}

For each of these 
QCCs, we also found the  minimum-length corresponding
tail-biting code that preserves minimum distance.  We first found the
minimum weight per cycle (``slope") $\alpha$ of the orthogonal code, and
then evaluated the upper bound $\Red{L} \le \lceil d^\perp/\alpha
\rceil$ of Handlery
\emph{et al.\ }\cite{HHJZ02} on the minimum number 
$\Red{L}$ of $3$-blocks
in the corresponding rate-$2/3$ tail-biting code. Using
\textsf{MAGMA} \cite{magma}, we then found the minimum distances
of tail-biting codes of up to this length to determine $\Red{L}$.

Table VII shows the minimum-length tail-biting codes corresponding to
the rate-$1/3$ binary convolutional codes of Table V. 

%\begin{figure}[hbt]
$$
\begin{array}{|r|c|cc|c|c|c|c|c|}
\hline
\rule[-7pt]{0pt}{20pt} %MG add some space for the fractions
\nu & d^\perp & \alpha & \lceil d^\perp/\alpha \rceil & N_{d^\perp} & 
\B & \B^\perp &
\mathrm{stabilizer~code} & d_{\mathrm{CSS}}
%& d_{\mathrm{opt}}
\\
\hline
  2&  3& 1/2  &  6& 2&(15, 5)  & (15, 10, 3) & [15, 5, 3]  & 3 %& 4
\\
\hline                                                   	
  3&  4& 1/2  &  8& 3&(21, 7)  & (21, 14, 4) & [21, 7, 4]  & 4
%& 5\mbox{--}6
\\
\hline                                                   	
  4&  4& 2/6  & 12& 1&(24, 8)  & (24, 16, 4) & [24, 8, 4]  & 4
%& 5\mbox{--}6
\\
  4&  4& 1/3  & 12& 1&(21, 7)  & (21, 14, 4) & [21, 7, 4]  & 4
%& 5\mbox{--}6
\\
  4&  4& 1/3  & 12& 1&(21, 7)  & (21, 14, 4) & [21, 7, 4]  & 4
%& 5\mbox{--}6
\\
\hline		
  5&  5& 1/3  & 15& 1&(39, 13) & (39, 26, 5) & [39, 13, 5] & 5\mbox{--}6
%& 7\mbox{--}10
\\
\hline                                                   	
  6&  6& 4/15 & 23& 2&(54, 18) & (54, 36, 6) & [54, 18, 6] & 6\mbox{--}8
%& 9\mbox{--}13
\\
\hline                                                   	
  7&  7& 5/18 & 26& 7&(63, 21) & (63, 42, 7) & [63, 21, 7] & \Red{7\mbox{--}10}
%& 10\mbox{--}
\\
\hline                                                   	
  8&  7& 3/11 & 26& 1&(69, 23) & (69, 46, 7) & [69, 23, 7] & \Red{8\mbox{--}10}
%& 10\mbox{--}
\\
  8&  7& 4/16 & 28& 1&(69, 23) & (69, 46, 7) & [69, 23, 7] & \Red{8\mbox{--}10}
%& 10\mbox{--}
\\
  8&  7& 3/14 & 33& 1&(60, 20) & (60, 40, 7) & [60, 20, 7] & \Red{7\mbox{--}9}
%& 9\mbox{--}
\\
  8&  7& 5/20 & 28& 1&(63, 21) & (63, 42, 7) & [63, 21, 7] & \Red{7\mbox{--}10}
%& 10\mbox{--}
\\
\hline                                                   	
  9&  8& 7/31 & 36& 3&(84, 28) & (84, 56, 8) & [84, 28, 8] & \Red{8\mbox{--}12}
%& 12\mbox{--}
\\
  9&  8&10/45 & 36& 3&(69, 23) & (69, 46, 8) & [69, 23, 8] & \Red{8\mbox{--}10}
%& 10\mbox{--}
\\
\hline                                                   	
10&  9& 9/41 & 41& 8&(99, 33) & (99, 66, 9) & [99, 33, 9] &\Red{9\mbox{--}14}
%& 12\mbox{--}
\\
\hline                                                   	
11&  9&11/52 & 43& 1&(105,35) &(105, 70, 9) &[105, 35, 9] &\Red{10\mbox{--}15}
%& 12\mbox{--}
\\
\hline\Red{
12& 10& 4/22 & 55& 5&(114,38) &(114, 76,10) &[114, 38,10] &10\mbox{--}\Red{16}
%& 14\mbox{--}	       
}\\
\hline
\end{array}
$$
\begin{center}
Table VII.  CSS-type rate-$1/3$ tail-biting codes.
\end{center}
%\end{figure}

Similarly, Table VIII shows the minimum-length tail-biting
codes corresponding to the rate-$1/3$ $\F_4$-linear convolutional codes
of Table VI.

%\begin{figure}[hbt]
$$
\begin{array}{|l|c|cc|c|c|c|c|c|}
\hline
\rule[-7pt]{0pt}{20pt} %MG add some space for the fractions
\nu & d^\perp & \alpha & \lceil d^\perp/\alpha \rceil & N_{d^\perp} & 
\B & \B^\perp &
\mathrm{stabilizer~code} & d_{\mathrm{opt}}\\
\hline
1& 3 & 1/1  &  3 &  3 & (9, 3)   & (9, 6, 3) & [9, 3, 3]   & 3\\
2& 4 & 2/3  &  6 & 12 & (15, 5)  &(15, 10, 4)& [15, 5, 4]  & 4\\
3& 5 & 1/3  & 15 &  3 & (24, 8)  &(24, 16, 5)& [24, 8, 5]  & 
5\mbox{--}6\\
4& 6 & 1/3  & 18 &  3 & (39, 13) &(39, 26, 6)& [39, 13, 6] & 
7\mbox{--}10\\
5& 8 & 6/14 & 19 & 75 & (45, 15) &(45, 30, 8)& [45, 15, 8] & 
8\mbox{--}11\\
\hline\Red{
6& 9 &  6/42  & 63 & 78 & (57,19) &(57,38,9)&[57,19,9]  & 9\mbox{--}14\\
6& 9 & 51/132 & 24 & 78 & (54,18) &(54,36,9)&[54,18,9]  & 9\mbox{--}13\\
6& 9 & 18/45  & 23 & 78 & (57,19) &(57,38,9)&[57,19,9]  & 9\mbox{--}14\\
6& 9 &  8/20  & 23 & 78 & (57,19) &(57,38,9)&[57,19,9]  & 9\mbox{--}14}\\
\hline
\end{array}
$$
\begin{center}
Table VIII.  $\F_4$-linear rate-$1/3$ tail-biting codes.
\end{center}
%\end{figure}

\section{Conclusion}

In this paper, we have introduced two types of quantum convolutional 
codes based on classical self-orthogonal rate-$1/n$ $\F_4$-linear and
$\F_2$-linear convolutional codes, respectively, with corresponding
decoders.  We have also introduced quantum tail-biting block codes 
based on these codes, which have the same rate, performance and decoding
complexity.  We have shown that these codes have a potentially 
attractive tradeoff between performance and complexity for
moderate-complexity applications.

In classical coding, convolutional coding was the next step beyond block
coding.  The next step was to concatenate convolutional codes with
algebraic (Reed-Solomon) outer codes for higher performance.  Finally, 
in the past decade, capacity-approaching codes such as low-density
parity-check (LDPC) codes and turbo codes with iterative decoding have
become the preferred techniques for highest performance.  One may
anticipate an analogous sequence of advances in quantum coding.  Indeed,
MacKay \etal have already taken a step toward quantum LDPC codes
\cite{MMM04}, although not without some difficulties.

\section*{Acknowledgments}

We wish to acknowledge helpful comments by Robert Calderbank, Emanuel
Knill and David MacKay.  Stefan H\"{o}st kindly provided a copy of
J\"{o}nsson's thesis \cite{J03}. M.~G. would like to thank Ingo
Boesnach for programming support. S.~G. wishes to acknowledge the
support of Prof.\ Jeffrey H. Shapiro and the U.S.  Army Research
Office (DoD MURI Grant No. DAAD-19-00-1-0177).

\pagebreak
%%%%%%%%%%%%%%%%%%%%%%%%%%%%%%%%

\end{document}